\documentclass[twocolumn]{aastex631}
\usepackage{newunicodechar}
\usepackage{amsmath}
\usepackage{color}

\DeclareUnicodeCharacter {2212} {-}
\DeclareUnicodeCharacter {02BC} {-}

\usepackage{txfonts}

\accepted{\today}
\submitjournal{ApJL}

\usepackage{graphicx}	
\usepackage{amsmath}	
\usepackage{color}
\usepackage{float}
\usepackage{soul}
\usepackage{multirow}

\def \n2dp{$\rm N_2D^+$}
\def \h13cop{$\rm H^{13}CO^+$}
\def \solarmass{$M_{\odot}$}
\def \dendro{\textit{Dendrogram}}
\def \imfit{\textit{CASA-imfit}}
\def \kms{km s$^{-1}$}

\begin{document}

\title{The ALMA-QUARKS survey: Detection of two extremely dense substructures in a massive prestellar core}
\shorttitle{ALMA-QUARKS: dense substructures inside a massive prestellar core}

\author[0000-0002-0786-7307]{Xiaofeng Mai}\thanks{E-mail:maixf@shao.ac.cn}
\affiliation{Shanghai Astronomical Observatory, Chinese Academy 
of Sciences, Shanghai 200030, People’s Republic of China}
\affiliation{School of Astronomy and Space Sciences, University of
Chinese Academy of Sciences, No. 19A Yuquan Road, Beijing 100049,
People’s Republic of China}

\author[0000-0002-5286-2564]{Tie Liu}\thanks{E-mail:liutie@shao.ac.cn}
\affiliation{Shanghai Astronomical Observatory, Chinese Academy 
of Sciences, Shanghai 200030, People’s Republic of China}

\author{Xunchuan Liu}
\affiliation{Shanghai Astronomical Observatory, Chinese Academy 
of Sciences, Shanghai 200030, People’s Republic of China}

\author{Lei Zhu}
\affiliation{Chinese Academy of Sciences South America Center for Astronomy, National Astronomical Observatories, Chinese Academy of Sciences, Beijing, 100101, PR China}

\author{Guido Garay}
\affiliation{Departamento de Astronom\'ia, Universidad de Chile, Las Condes, 7591245 Santiago, Chile}

\author{Paul F. Goldsmith}
\affiliation{Jet Propulsion Laboratory, California Institute of Technology, 4800 Oak Grove Drive, Pasadena, CA 91109, USA}

\author[0000-0002-5809-4834]{Mika Juvela}
\affiliation{Department of Physics, P.O. box 64, FI- 00014, University of Helsinki, Finland}

\author[0000-0003-3343-9645]{Hongli Liu}
\affiliation{Department of Astronomy, Yunnan University, Kunming 650091, People’s Republic of China}

\author[0000-0002-0402-3029]{Emma Mannfors}
\affiliation{Department of Physics, P.O. box 64, FI- 00014, University of Helsinki, Finland}

\author[0000-0001-5917-5751]{Anandmayee Tej}
\affiliation{Indian Institute of Space Science and Technology, Thiruvananthapuram 695 547, India}

\author[0000-0002-7125-7685]{Patricio Sanhueza}
\affiliation{National Astronomical Observatory of Japan, National Institutes of Natural Sciences, 2-21-1 Osawa, Mitaka, Tokyo 181-8588, Japan}
\affiliation{Astronomical Science Program, The Graduate University for Advanced Studies, SOKENDAI, 2-21-1 Osawa, Mitaka, Tokyo 181-8588, Japan}

\author[0000-0003-1275-5251]{Shanghuo Li}
\affiliation{Max Planck Institute for Astronomy, Königstuhl 17, D-69117 Heidelberg, Germany}

\author[0000-0001-5950-1932]{Fengwei Xu}
\affiliation{Kavli Institute for Astronomy and Astrophysics, Peking University, 5 Yiheyuan Road, Haidian District, Beijing 100871, Peopleʼs Republic of China}

\author[0000-0002-1424-3543]{Enrique Vazquez Semadeni}
\affiliation{Instituto de Radioastronom\'ia y Astrof\'isica, Universidad Nacional Aut\'onoma de M\'exico, Antigua Carretera a P\'atzcuaro 8701, Ex-Hda. San Jos\'e de la Huerta, 58089 Morelia, Michoac\'an, M\'exico}

\author{Wenyu Jiao}
\affiliation{Kavli Institute for Astronomy and Astrophysics, Peking University, 5 Yiheyuan Road, Haidian District, Beijing 100871, Peopleʼs Republic of China}
\affiliation{Department of Astronomy, School of Physics, Peking University, Beijing, 100871, People's Republic of China}
\affiliation{Max Planck Institute for Astronomy, Königstuhl 17, D-69117 Heidelberg, Germany}

\author{Yaping Peng}
\affiliation{Department of Physics, Faculty of Science, Kunming University of Science and Technology, Kunming 650500, People’s Republic of China}

\author[0000-0003-0295-6586]{T. Baug}
\affiliation{S. N. Bose National Centre for Basic Sciences, Block JD, Sector III, Salt Lake, Kolkata 700106, India}

\author[0000-0003-4546-2623]{Aiyuan Yang}
\affiliation{National Astronomical Observatories, Chinese Academy of Sciences, A20 Datun Road, Chaoyang District, Beijing, 100101, P. R. China}

\author[0000-0001-6725-0483]{Lokesh Dewangan}
\affiliation{Physical Research Laboratory, Navrangpura, Ahmedabad 380009, India }

\author[0000-0002-9574-8454]{Leonardo Bronfman}
\affiliation{Departamento de Astronom\'a, Universidad de Chile, Las Condes, 7591245 Santiago, Chile}

\author[0000-0003-4714-0636]{Gilberto C. G\'omez}
\affiliation{Instituto de Radioastronom\'ia y Astrof\'isica, Universidad Nacional Aut\'onoma de M\'exico, Antigua Carretera a P\'atzcuaro 8701, Ex-Hda. San Jos\'e de la Huerta, 58089 Morelia, Michoac\'an, M\'exico}

\author[0000-0002-9569-9234]{Aina Palau}
\affiliation{Instituto de Radioastronom\'ia y Astrof\'isica, Universidad Nacional Aut\'onoma de M\'exico, Antigua Carretera a P\'atzcuaro 8701, Ex-Hda. San Jos\'e de la Huerta, 58089 Morelia, Michoac\'an, M\'exico}

\author{Chang Won Lee}
\affiliation{Korea Astronomy and Space Science Institute, 776 Daedeokdae-ro, Yuseong-gu, Daejeon 34055, Republic of Korea}
\affiliation{University of Science and Technology, Korea (UST), 217 Gajeong-ro, Yuseong-gu, Daejeon 34113, Republic of Korea}

\author[0000-0003-2302-0613]{Sheng-Li Qin}
\affiliation{Department of Astronomy, Yunnan University, Kunming 650091, People’s Republic of China}

\author[0000-0002-8149-8546]{Ken'ichi Tatematsu}
\affiliation{Nobeyama Radio Observatory, National Astronomical Observatory of Japan, National Institutes of Natural Sciences, 462-2 Nobeyama, Minamimaki, Minamisaku, Nagano 384-1305, Japan}
\affiliation{Astronomical Science Program, The Graduate University for Advanced Studies, SOKENDAI, 2-21-1 Osawa, Mitaka, Tokyo 181-8588, Japan}

\author[0000-0002-9875-7436]{James O. Chibueze}
\affiliation{Department of Mathematical Sciences, University of South Africa, Cnr Christian de Wet Rd and Pioneer Avenue, Florida Park, 1709, Roodepoort, South Africa}
\affiliation{Department of Physics and Astronomy, Faculty of Physical Sciences, University of Nigeria, Carver Building, 1 University Road, Nsukka 410001, Nigeria}

\author{Dongting Yang}
\affiliation{Department of Astronomy, Yunnan University, Kunming 650091, People’s Republic of China}

\author{Xing Lu}
\affiliation{Shanghai Astronomical Observatory, Chinese Academy 
of Sciences, Shanghai 200030, People’s Republic of China}

\author[0000-0003-4506-3171]{Qiuyi Luo}
\affiliation{Shanghai Astronomical Observatory, Chinese Academy 
of Sciences, Shanghai 200030, People’s Republic of China}
\affiliation{School of Astronomy and Space Sciences, University of
Chinese Academy of Sciences, No. 19A Yuquan Road, Beijing 100049,
People’s Republic of China}

\author{Qilao Gu}
\affiliation{Shanghai Astronomical Observatory, Chinese Academy 
of Sciences, Shanghai 200030, People’s Republic of China}

\author[0000-0002-7881-689X]{Namitha Issac}
\affiliation{Shanghai Astronomical Observatory, Chinese Academy 
of Sciences, Shanghai 200030, People’s Republic of China}

\author[0000-0002-8389-6695]{Suinan Zhang}
\affiliation{Shanghai Astronomical Observatory, Chinese Academy 
of Sciences, Shanghai 200030, People’s Republic of China}

\author[0000-0001-8077-7095]{Pak-Shing Li}
\affiliation{Shanghai Astronomical Observatory, Chinese Academy 
of Sciences, Shanghai 200030, People’s Republic of China}

\author{Bo Zhang}\thanks{E-mail:zb@shao.ac.cn}
\affiliation{Shanghai Astronomical Observatory, Chinese Academy 
of Sciences, Shanghai 200030, People’s Republic of China}

\author[0000-0002-5310-4212]{L. Viktor T\'oth}
\affiliation{Institute of Physics and Astronomy, E\"otv\"os Lor\`and University, P\'azm\'any P\'eter s\'et\'any 1/A, H-1117 Budapest, Hungary}

\begin{abstract}

Only a handful of massive starless core candidates have been discovered so far, but none of them have been fully confirmed. Within the MM1 clump in the filamentary infrared dark cloud G34.43+0.24 that was covered by
the ALMA-ATOMS survey at Band 3 ($\sim2\arcsec$, 6000\,au) and the ALMA-QUARKS survey at Band 6 ($\sim 0.3\arcsec$, 900\,au), two prestellar core candidates MM1-C and E1 with masses of 71 and 20 \solarmass~and radii of 2100--4400\,au were discovered.
The two cores show no obvious sign of star-formation activities. In particular, MM1-C is a very promising massive prestellar core candidate with a total gas mass of 71\,\solarmass. Within MM1-C, we detected two extremely dense substructures, C1 and C2, as characterized by their high densities of $\rm n_{H_2}\sim 10^{8-9} cm^{-3}$. Moreover, evidence of further fragmentation in C2 was also revealed. We have detected the primordial fragmentation in the earliest stage of massive star formation, and we speculate that MM1-C would be the birthplace of a massive multiple system. However, we cannot fully rule out the possibility that the massive prestellar core MM1-C will just form a cluster of low-mass stars if it undergoes further fragmentation.

\end{abstract}


\section{Introduction}
Massive stars with their enormous UV radiation and large mechanical energy output, profoundly impact the surrounding interstellar medium (ISM) and play a pivotal role in the evolution of galaxies and even the Universe. However, observing the progenitors of massive stars is very challenging due to their rapid evolutionary timescale and the highly opaque molecular clouds surrounding them. Consequently, the formation of massive stars remains enigmatic, which has led to very diverse ideas on the mechanisms involved. 
For instance, the \textit{Turbulent Core} model \citep[e.g.][]{2003ApJ...585..850M} proposes that massive stars form in the same way as low-mass stars, by gradually accumulating gas from their natal cores (sizes $\sim$ 0.1\,pc). In this case, massive prestellar cores should thus be the progenitors of massive stars. 
Conversely, the ``clump-feed" models such as \textit{Competitive Accretion} \citep[e.g.][]{2001MNRAS.323..785B}, the \textit{Inertial Inflow} \citep[e.g.][]{2020ApJ...900...82P}, and the \textit{Global Hierarchical Collapse} model \citep{2009ApJ...707.1023V, 2019MNRAS.490.3061V} suggest that massive stars form from low-mass cores within clouds amassing through further gas accretion from their natal clumps or even clouds.

Hunting for massive prestellar cores is one of the most straightforward ways to distinguish among different scenarios. Due to their rapid evolution, only a few promising massive prestellar core candidates have been identified to date: e.g. 
\citet[core C2c1a in the dragon cloud]{2023A&A...675A..53B}, \citet[W43-MM1\#6]{2018A&A...618L...5N}, and \citet[G9.62+0.19MM9]{Liu2017}, etc. 
However, none of these massive starless core candidates have been fully confirmed, indicating that the search for massive prestellar core candidates warrants further observational efforts.
Given that the current sample size is small, a comprehensive understanding requires persistent efforts in identifying additional samples of massive prestellar core candidates. 
On the other hand, surveys aiming to identify massive prestellar cores in the earliest phases of massive clumps have yielded no conclusive results in their favor \citep{2017ApJ...841...97S, 2019ApJ...886..102S, 2021ApJ...923..147M, 2023ApJ...950..148M, 2023ApJ...949..109L}. Specifically, \citet{2023ApJ...950..148M} reported the absence of massive prestellar cores within a substantial sample of 70\,$\rm \mu m$ dark IRDCs, encompassing over 800 cores. \citet{2019ApJ...886..102S} suggested that if massive prestellar cores do indeed exist, they are likely embedded in the more evolved environment.
Recently, in an ALMA 0.87\,mm survey of 11  clumps, \citet{Xu2023ASSEMBLE} confirmed that starless cores are becoming more massive in evolved clumps as compared to early-stage clumps, through continuous gas accretion. Therefore, it might be more promising to search for massive starless cores in a protocluster environment.

Located 3.0\,kpc away from the Sun \citep{2023ApJ...949...10M}, the filamentary IRDC G34.43+0.24 (hereafter G34) is a well studied high-mass star forming region \citep{2004ApJ...610..313G, 2005ApJ...630L.181R, 2010ApJ...715...18S, 2012ApJ...756...60S, 2013ApJ...775L..31S, 2014ApJ...791..108F, 2016ApJ...819..117X, 2019ApJ...883...95S, 2020ApJ...901...31L}. Molecular outflows have been reported for the dense core G34-MM1 by \citet{2010ApJ...715...18S} and \citet{2018ApJS..235....3Y} at low resolution ($\sim 18\arcsec$). Given that molecular outflows are one of the earliest signatures of star formation activities \citep{2022A&A...658A.160Y}, the dense core G34-MM1 in the IRDC G34 is supposed to be in the early stage of star formation.  
\citet{2022MNRAS.510.5009L} have extensively studied this complex using the high angular resolution ($\sim 2\arcsec$) data of the ATOMS survey \citep{2020MNRAS.496.2790L},
and identified six dense cores in its G34-MM1 clump with masses ranging from 20 to 122\,\solarmass \footnote{The mass in \citet{2022MNRAS.510.5009L} was estimated based on the kinematic distance \citep{2001ApJ...551..747S, 2006ApJ...653.1325S}. Those values in this paper are corrected using the parallax distance in \citet{2023ApJ...949...10M}.}.
However, the evolutionary stages of these cores need to be explored further.
The ALMA-QUARKS survey \citep[][see also Section \ref{sec:obs}]{liu2023almaquarks} observed the densest kernels of the ATOMS sources at Band 6 with a much higher resolution of $\sim 0.3\arcsec$, providing the possibility
of verifying the existence of massive starless cores in G34-MM1 in a spatial resolution of $\sim 1000$ au.
In this letter, we revisit these cores and report the discovery of the two cores MM1-C and MM1-E1,  situated close to the boundary of G34-MM1 clump, as candidate massive prestellar cores containing extremely dense substructures.  

\begin{figure*}
    \centering
    \includegraphics[width=0.85\linewidth]{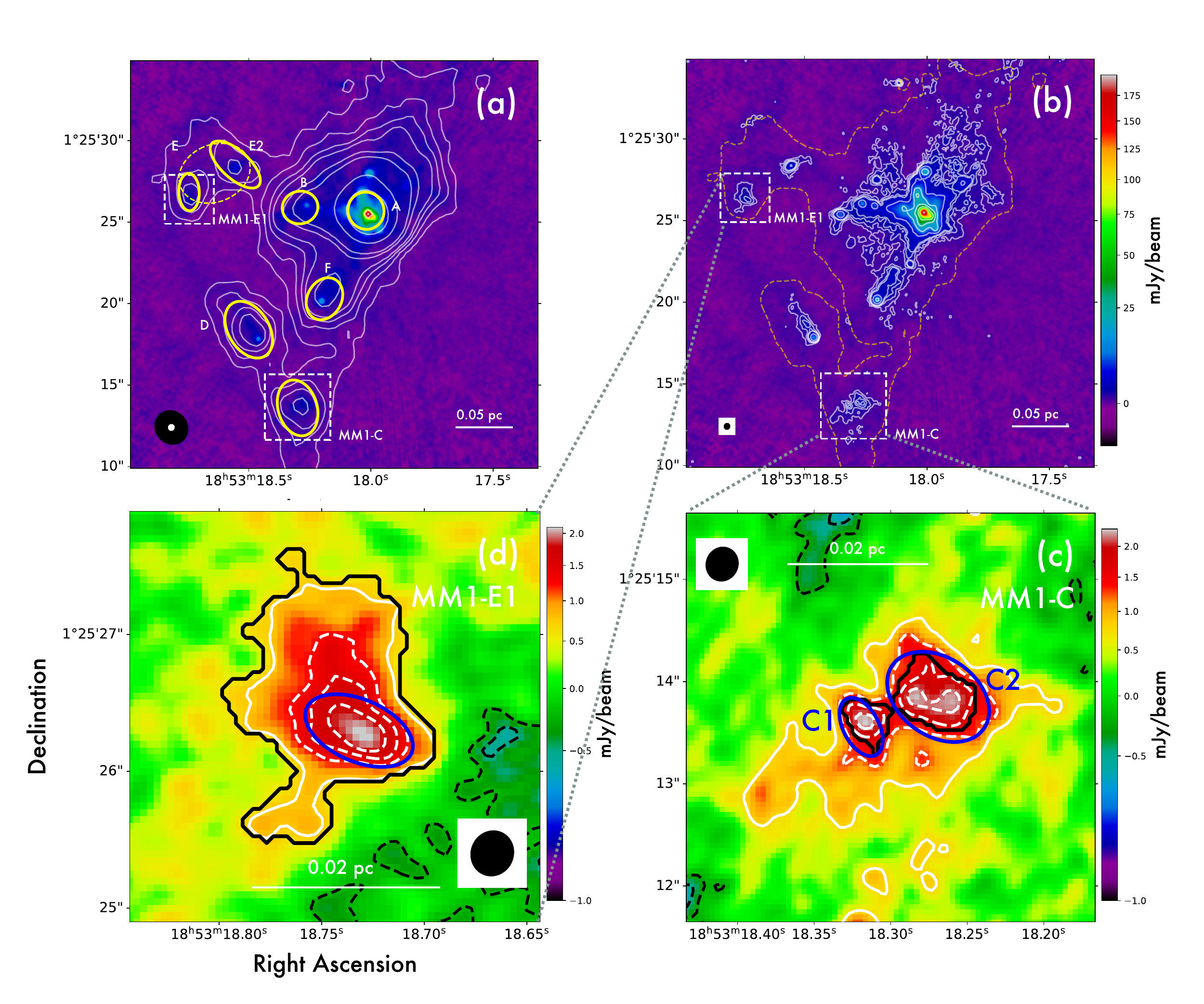}
    \caption{ALMA Band 3 and 6 continuum emission of G34.43+0.24 MM1 and the detected prestellar cores MM1-C and MM1-E1. Panel (a) shows Band 6 continuum emission overplotted with Band 3 continuum emission~\citep{2022MNRAS.510.5009L} using white contours of [3, 5, 7, 9, 15, 20, 50, 100, 200] $\times~\sigma_{\rm b3}$, where $\rm \sigma_{\rm b3}=0.3~mJy~beam^{-1}$. The yellow solid ellipses labeled A-D and F and the dashed ellipse labeled E identify the cores extracted in \citet{2022MNRAS.510.5009L}. The solid ellipses labeled E1 and E2 represent the re-estimated FWHM sizes of core E1 and E2.
    Panel (b) shows the contours of Band 6 continuum emission overplotted on the same colormap as Panel (a). The white contours are at levels of [3, 5, 7, 9, 20, 50, 100, 200] $\times~\sigma_{\rm b6}$, where $\rm \sigma_{\rm b6}=0.25~mJy~beam^{-1}$. The orange dashed contour outlines the 3-$\sigma_{\rm b3}$ boundary of Band 3 continuum emission.
    Panel (c) zooms toward core MM1-C. The contours are at [3, 5, 7, 9] $\times~\sigma_{\rm b6}$. The solid contour highlights the 3-$\sigma_{\rm b6}$ boundary. 
    Panel (d) zooms toward core MM1-E1 with contours at [3, 5, 7]$\times~\sigma_{\rm b6}$. The solid contour also emphasizes the 3-$\sigma_{\rm b6}$ boundary.
    The colorbars of both Panel (c) and (d) are set to start from 0 and end at their peak intensities to display the morphology of MM1-C and MM1-E1 more clearly.
    The blue ellipses in Panels (c) and (d) correspond to the FWHM sizes estimated from the 2-dimensional Gaussian fitting.
    The restoring beams are shown as black ellipses in each panel.
    The black solid contours outline the condensations identified by dendrogram.
    The black dashed contour is at [-2, -1]$\times\sigma_{\rm b6}$.
    }
    \label{fig:continuum}
\end{figure*}

\section{Observations}
\label{sec:obs}
G34-MM1 was observed as part of the ALMA project ``ALMA Three-millimeter Observations of Massive Star-forming regions" (ATOMS, Project ID: 2019.1.00685.S; PI: Tie Liu). The observations were conducted with the 7-m array (ACA) and 12-m arrays (C43-2 or C43-3) at Band 3. The observation setup and data reduction can be found in \citet{2020MNRAS.496.2790L}. The angular resolution and maximum recoverable scales (MRS) for the ACA and 12-m array are 13\farcs3 and 76\farcs9, and 1\farcs5 and 19\farcs4, respectively. The ACA and 12-m combined data have resulted in a resolution of $\sim2\arcsec$ for continuum emission. The combined continuum image of G34-MM1 reaches an rms noise level of 0.3\,mJy~beam$^{-1}$. Eight spectral windows (SPWs) were set including six SPWs with high spectral resolution and two wide SPWs with 1875 MHz bandwidth. The spectral data reaches a noise level of 12\,mJy beam$^{-1}$ per 0.062 MHz channel ($\sim$0.2--0.4 \kms) for six SPWs, and 3.6\,mJy beam$^{-1}$ per 0.488 MHz channel ($\sim 1.6$ \kms) for wide SPWs.

G34-MM1 was also observed as part of a new ALMA project, ``Querying Underlying mechanisms of massive star formation with ALMA-Resolved gas Kinematics and Structures" (QUARKS, PIs: Lei Zhu, Guido Garay and Tie Liu). The QUARKS project targets are the same objects as in the ATOMS survey but with a much higher resolution, of $\sim0.25\arcsec$, at Band 6 (ALMA Project IDs: 2021.1.00095.S and 2023.1.00425.S; PI: Lei Zhu). The observations were conducted with ACA and 12-m arrays (C43-2 and C43-5) at Band 6. The observation setup and data reduction of QUARKS are described in \citet{liu2023almaquarks}. The ACA and 12-m combined data were resulting in an angular resolution of 0\farcs3, and the corresponding combined continuum image achieves a rms noise level of 0.25\,mJy beam$^{-1}$. Four wide SPWs with 1875 MHz bandwidth were set, resulting in a noise level of 5\,mJy beam$^{-1}$ per 0.488 MHz channel ($\sim 0.6$ \kms).

\begin{figure*}
    \centering
    \includegraphics[width=0.95\linewidth]{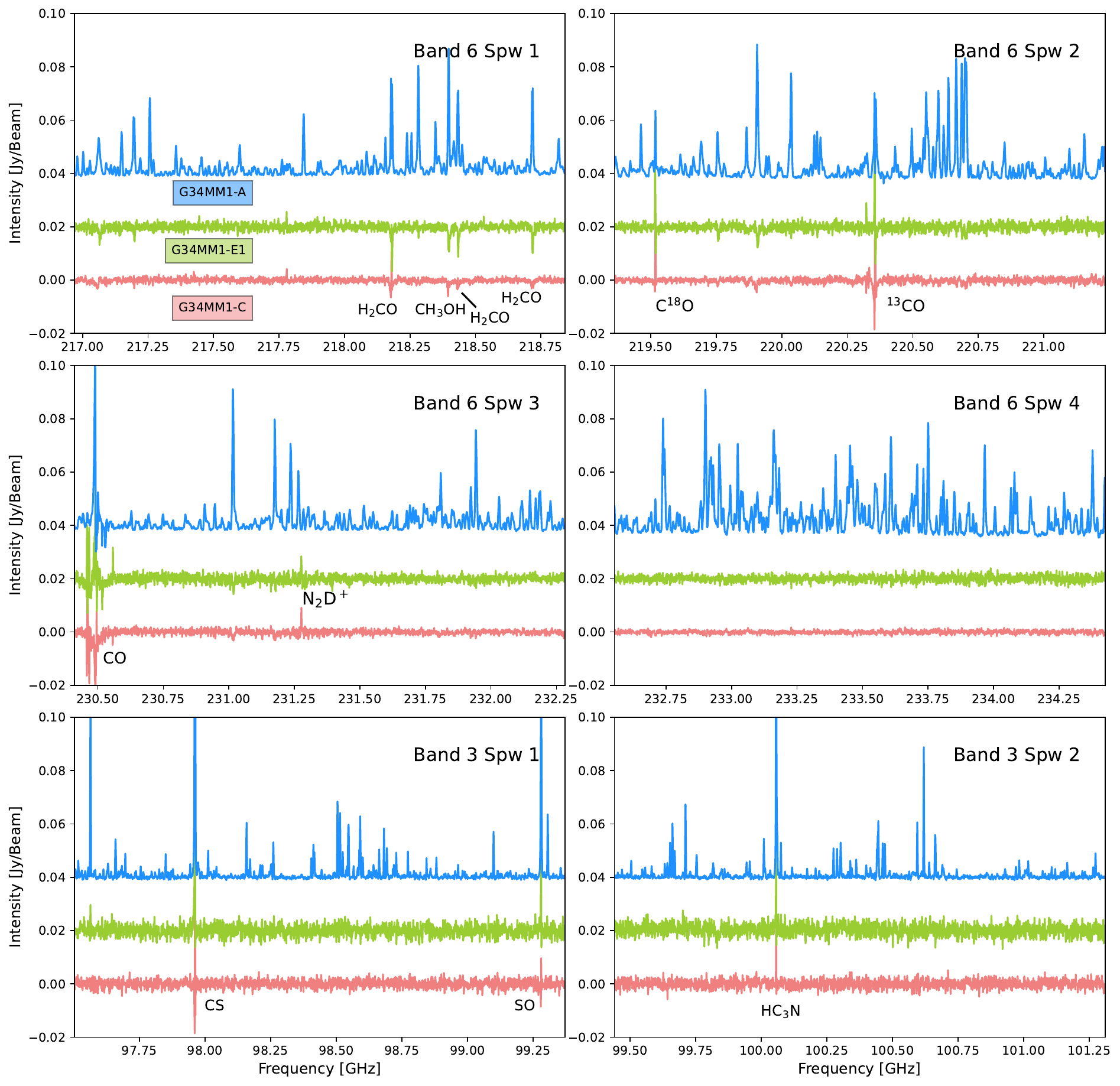}
    \caption{Core-averaged spectra of all four windows at Band 6 and two windows at Band 3 at the positions of two candidate prestellar cores MM1-C (red) and MM1-E1 (green), and the hot core MM1-A(blue) for comparison \citep{2022MNRAS.511.3463Q}. The spectra are averaged over a circular area of $2\arcsec.5$ in diameter (shown as green circles in Figure \ref{fig:appendixa}). The offset of intensity for MM1-E1 and MM1-A is for visual clarity. The intensity of the spectra from MM1-A was divided by 5 for Spw 1, 2, 3 at Band 6, and by 4 for Spw 1, 2 at Band 3, and Spw 4 at Band 6. The spectra show the line-poor characteristics of MM1-C and MM1-E1 indicating their possible prestellar nature, and the clear detection of \n2dp indicating the cold environment. Appendix~\ref{sec:appb} also shows the beam-averaged spectra.
    }
    \label{fig:spectra}
\end{figure*}

\begin{table*}
  \centering
  \caption{Core properties.}
  \label{tab:coreproperties}
  \begin{tabular}{lccccccccc}
    \hline\hline
Source & R.A.        & Decl.      & $\rm FWHM^{Deconv}$      & PA        & $R_{\rm eff}$      & $S_{\nu}$ & $T_{\rm gas}$ & $M_{\rm gas}$ & $n_{\rm H_2}$                   \\
       & (J2000)     & (J2000)    & (\arcsec) & ($\degr$) & (au) & (mJy)     & (K)       & ($M_{\odot}$) & ($\rm 10^7~cm^{-3}$)  \\
\hline
   \large\textit{ Band 3$^*$} &             &            &                          &           &                  &           &           &               &                                           \\
MM1-C  & 18:53:18.30 & 1:25:13.60 & $3.5\times2.4$           & 16.0      & $4390\pm 239$                 & $9.1$       & 14.2      & $71.2\pm33.4$            & $2.6\pm1.3$                  \\
MM1-E1 & 18:53:18.74 & 1:25:26.84 & $2.3\times1.3$           & 2.1       & $2169\pm 142$                 & $4.2$       & 22.6      & $20.3\pm9.6 $            & $3.5\pm1.8$                      \\
\hline
  {\large\textit{ Band 6}}  &             &            &                          &           &                   &           &           &               &                                        \\
MM1-C1 & 18:53:18.32 & 1:25:13.56 & $0.62\times0.35$         & 27.9      & $701\pm 38$                & $6.1$       & 14.2      & $2.1\pm1.1$          & $18.4\pm11.2$                           \\
       &             &            &                          &           &                    &           & 10        & $3.7\pm1.7$          & $32.7\pm16.0$                                     \\
       &             &            &                          &           &                    &           & 6.5       & $8.2\pm3.8$          & $71.7\pm35.9$                                     \\
MM1-C2 & 18:53:18.27 & 1:25:13.85 & $1.09\times0.77$         & 55.6      & $1387\pm 75$                & $21.2$      & 14.2      & $7.3\pm3.4$          & $8.4\pm4.4$                            \\
       &             &            &                          &           &                    &           & 10        & $12.9\pm6.0$         & $14.9\pm7.5$                                      \\
       &             &            &                          &           &                    &           & 6.5       & $28.4\pm13.2$         & $32.8\pm16.3$                                    \\
MM1-E1 & 18:53:18.73 & 1:25:26.30 & $0.83\times0.45$         & 66.8      & $925\pm 50$                & $10.2$      & 22.6      & $2.0\pm0.9$          & $7.6\pm3.8$                             \\
    \hline
  \end{tabular}
  \tablecomments{ *: The coordinate, position angle PA, and integrated flux of MM1-C are taken from \citet{2022MNRAS.510.5009L}, the rest of the parameters are recalculated based on the parallax distance. The uncertainties for $R_{\rm eff}$, $M_{\rm gas}$, and $n_{\rm H_2}$ are the propagation from the uncertainty in the parallax distance, opacity, flux, and gas-dust ratio.}
 \end{table*}

\section{Results}
\label{sec:results}
\subsection{Detection of massive starless core candidates }
\label{sec:3.1}
Fig.~\ref{fig:continuum} presents the ALMA Band 3 (3\,mm) and Band 6 (1.3\,mm) continuum emission maps of the G34-MM1 clump. Reexamination of the Band 3 continuum emission revealed that the MM1-E core identified in \citet{2022MNRAS.510.5009L} is in fact two separate cores (namely E1, E2) instead of one single core, as clearly seen from higher-resolution 1.3\,mm emission. We performed two-component two-dimensional Gaussian fitting toward E1 and E2 following the same procedure as in \citet{2021MNRAS.505.2801L}, which is a combination of \dendro~\citep{2008ApJ...679.1338R}~and \imfit~for consistency. The measured core properties of these two cores are summarized in Table~\ref{tab:coreproperties}. In this work, we only focus on the two massive starless core candidates MM1-C and MM1-E1.

Since the MRS of the Band 3 data is two times larger than that of the Band 6 data and the 3\,mm continuum emission is relatively optically thin than the 1.3\,mm continuum emission, we use the 3\,mm continuum emission to calculate the total gas masses of dense cores MM1-C and MM1-E1. Assuming that the 3\,mm continuum emission is optically thin under local thermodynamic equilibrium (LTE) conditions,  the core masses can be calculated as \citep[e.g.,][]{2021ApJ...912..148L},
\begin{equation}
    M_{\rm gas}=\frac{R_{\rm gd}S_{\nu}D^2}{\kappa_{\nu}B_{\nu}(T_{\rm dust})},
    \label{eqa:mass}
\end{equation}
where $S_{\nu}$ is the integrated 3\,mm continuum flux, $D$ is the distance to the region (3.0\,kpc), and $B_{\nu}(T_{\rm dust})$ is the Planck function at dust temperature $T_{\rm dust}$. The dust temperature $T_{\rm dust}=$14.2\,K and 22.6\,K for MM1-C and E1 respectively are taken as the gas kinetic temperature of $\rm NH_3$ from VLA observations at the resolution of $\sim$ 3\arcsec \citep{2014ApJ...790...84L}. We assume a typical gas-dust mass ratio $R_{\rm gd}$ of 100. A value for $\kappa_\nu$ of 0.18 $\rm cm^{2}~g^{-1}$ is adopted from the `OH5' model\citep{1994A&A...291..943O, 1994ApJ...421..615P} and extended to the longer wavelength by \citet{2005ApJ...627..293Y}.
Furthermore, by assuming a spherical geometry, the number density can be estimated as $n_{\rm H_2}=3M/(4\pi\mu_{\rm H_2}m_{\rm H}R_{\rm core}^3)$, where the molecular weight per hydrogen molecule $\mu_{\rm H_2}$ is 2.8, and $m_{\rm H}$ is the mass of a proton. The effective radii of the cores $R_{\rm core}$ are estimated as 
$\rm (FWHM^{Deconv}_{major}\times FWHM^{Deconv}_{minor})^{\frac{1}{2}}/2\times D$ 
where $\rm FWHM^{Deconv}_{major}$ and $\rm FWHM^{Deconv}_{minor}$ are the deconvolved FWHM of the major and minor axis, respectively.

Several sources of uncertainty can influence the mass determination. Following \citet{2017ApJ...841...97S, 2019ApJ...886..102S}, we conservatively assign the uncertainties of 25\% for $R_{\rm dg}$ and $\kappa_\nu$, 5\% for the absolute flux uncertainty at Band 3\footnote{ALMA proposal guide:\url{https://almascience.eso.org/proposing/proposers-guide##autotoc-item-autotoc-79}} and the distance ($3.03^{+0.17}_{-0.16}$\,kpc). The combined uncertainty of mass determination is $\sim$ 50\%.
The mass, deconvolved size, number density, and column density estimated from Band 3 continuum emission are summarized in Table~\ref{tab:coreproperties}. We find that the cores MM1-C and E1 are massive and dense, having masses of 71\,\solarmass~and 20\,\solarmass~, respectively, on a scale of approximately 4000--8000\,au (0.02--0.04\,pc). Their densities reach up to $\sim3\times10^7$ cm$^{-3}$. Considering a star formation efficiency (a conversion efficiency of core
mass into star mass; $\epsilon_{\rm core}$) of 50\%, which is reasonable and maybe even conservative for extreme density cores \citep{2010A&A...524A..18B, 2013ApJ...762..120P, Motte2018,Pouteau2022}, the two massive and dense cores in G34-MM1 clump (MM1-C and E1) will have the potential to form a massive star or a small star cluster
with a stellar mass $M_{\rm star}>$8 $M_\sun$ eventually. 

Fig.~\ref{fig:spectra} shows the core-averaged spectra of MM1-C and E1 in four wide-band spectral windows at Band 6, and two wide-band spectral windows at Band 3. For comparison, we also plot the spectra of the hot molecular core MM1-A, which has already formed high-mass protostars \citep{2022MNRAS.511.3463Q}. It can be easily noticed that very few lines are detected toward MM1-C and E1, which is in strong contrast with the `line forest' of MM1-A. Some absorption features in the spectra of molecules including H$_2$CO, $^{13}$CO, C$^{18}$O, and CH$_3$OH are likely caused by the missing flux issue as a result of interferometer observations due to sidelobe or a lack of shorter baselines and total power (TP) data. As shown in the moment 0 maps in Fig. \ref{fig:appendixa}, no clear emission or absorption features of CH$_3$OH with signal-to-noise ratios (SNRs) larger than 3$\sigma$ were detected toward both MM1-C and E1. In addition, as shown in Figure \ref{fig:appendixb}, the beam-averaged spectra at band 6 toward the continuum emission peak of MM1-C1 show no absorption features in either H$_2$CO or CH$_3$OH lines, which could be caused by warm or hot continuum background sources (or protostars) within MM1-C1. 

Additionally, \citet{2022MNRAS.510.5009L} identified six outflow lobes in the MM1 clump, but none of them seem to come from MM1-C or E1. We also carefully checked the $^{12}$CO \textit{J}=2-1 and SiO \textit{J}=5-4 lines toward the two cores and found no evidence of high-velocity emission. We present the CO outflows in Fig. \ref{fig:appendixc}. Both MM1-C and E1 are far away from CO outflow lobes.
Also, no radio jets are detected by the GLOSTAR survey \citep{2023A&A...670A...9D} associated with these two cores \citep{2023arXiv231009777Y}.
Taking into account both the paucity of spectral lines and the absence of outflows, we postulate that there is no ongoing star-formation activity within cores MM1-C and E1. Thus, we classify these two cores as starless cores, among which MM1-C is a more promising massive starless core \citep[$>30\,M_{\odot}$, e.g.,] []{2017ApJ...841...97S}.\\

\begin{table}
\centering
\small
\caption{Jeans and virial analysis.}
\begin{tabular}{l|l|ccc|ccc}
\hline\hline
\multirow{3}{*}{Source} &  & \multicolumn{3}{l|}{$\sigma_{\rm v, N_2D^+}=0.74~{\rm km~s^{-1}}$} & \multicolumn{3}{l}{$\sigma_{\rm v, H^{13}CO^+}=0.46~{\rm km~s^{-1}}$} \\ \cline{3-8} 
 & $T_{\rm gas}$ & $L_{\rm J}$ & $M_{\rm J}$ & $\alpha$ & $L_{\rm J}$ & $M_{\rm J}$ & $\alpha$ \\
 & (K) & (au) & ($M_{\odot}$) &  & (au) & ($M_{\odot}$) &  \\ \hline
 MM1-C & 14.2 & -- & -- & 0.76 & -- & -- & 0.28 \\ \hline
\multirow{3}{*}{MM1-C1} & 14.2 & 2146 & 7.25 & 3.86 & 1276 & 1.52 & 1.69 \\
 & 10 & 1685 &  5.69 & 2.13 & 1002 & 1.20 & 0.91 \\
 & 6.5 & 1139 & 3.85 & 0.95 & 677 & 0.81 & 0.40 \\ \hline
\multirow{3}{*}{MM1-C2} & 14.2 & 3248 & 10.97 & 2.06 & 1931 & 2.31 & 0.90 \\
 & 10 & 2521 & 8.52 & 1.21 & 1499 & 1.79 & 0.52 \\
 & 6.5 & 1704 & 5.76 & 0.54 & 1013 & 1.21 & 0.23 \\ \hline
\end{tabular}
  \tablecomments{The effective diameters of MM1-C1 and C2 are 1400 and 2760 au, respectively.}
\label{tab:dynamics}
\end{table}

\subsection{Detection of extremely dense Substructures within Core MM1-C}
\label{sec:3.2}

Panels (c) and (d) in Fig.~\ref{fig:continuum} show high-resolution Band 6 (1.3 mm) continuum emission at a resolution of $\sim$0\farcs 3 toward the two massive cores MM1-C and MM1-E. MM1-E1 shows no evidence of fragmentation, remaining a single condensation at this resolution, whereas two prominent condensations (namely, C1 and C2) are clearly seen in MM1-C.
The fragmentation could be interferometric artifacts, as discussed in \citet{2019ApJ...874...89C}. 
However, the detection was at the 3$\sigma$ level in their synthetic observations, while in our observations, the detection of the fragments is more reliable, at $\sim10\sigma$ level. In particular, the fragments are well separated by 2400 au (about 3 beam sizes, or 0\farcs8 at 3\,kpc), and can be reliably identified by the \dendro~algorithm. Therefore, we argue that core MM1-C shows obvious fragmentation.

As shown by the black contour in Panel (c) in Fig.~\ref{fig:continuum}, two condensations were identified by \dendro\footnote{An intensity threshold of 2.5$\sigma_{\rm b6}$, step of 1$\sigma_{\rm b6}$, and a minimum pixel number contained in a restoring beam, 44 in this case, were fed into the \dendro~algorithm.}. Due to the proximity, we performed a simultaneous 2D Gaussian fitting of the cores. The estimated sizes of C1 and C2 are 701 and 1381 au, respectively. We also assign the uncertainties of the parameters in Eq.~(\ref{eqa:mass}) as in Sec.~\ref{sec:3.1}, with the exception of the absolute flux, which has an uncertainty of 10\% at Band 6 (See also ALMA proposal guide). For $\kappa_\nu$ at Band 6, we adopt a value of 0.9\,$\rm cm^2\,g^{-1}$ \citep{1994A&A...291..943O}.
Assuming they have the same temperature as the natal core, the masses of C1 and C2 are $\sim$2.1\,\solarmass ~and $\sim$7.3\,\solarmass, respectively. The derived densities are remarkably high, of the order $\rm \sim10^8\,cm^{-3}$. Negative temperature gradients of prestellar cores have been reported \citep{2007A&A...470..221C, 2021ApJ...923..231S}.
The core-average temperature is possibly an upper limit, and therefore the estimated masses are lower limits. We further calculate their masses assuming lower temperatures of 10\,K and 6.5\,K. The corresponding parameters of C1 and C2 are also listed in Table~\ref{tab:coreproperties}.
The mass of MM1-C2 can be as high as 10-30\,$M_\sun$ if $T_{\rm dust}\leq 10$\,K. In addition, MM1-C2 could further accumulate its mass through accreting gas from its collapsing natal core as well as a cold filament (see section \ref{sec:n2dp}). Within such a small-scale region ($\sim1000$ au) containing extremely high density ($\sim10^8$ cm$^{-3}$) gas, a higher $\epsilon_{\rm core}\sim50\%-100\%$ could be possible \citep{2010A&A...524A..18B, 2013ApJ...762..120P, Benedettini2018,Motte2018,Pouteau2022}. Thus, it is possible that MM1-C2 can form a massive star ($M_{\rm star}>$8 $M_\sun$) in the future.

\begin{figure*}[!thb]
    \centering
    \includegraphics[width=18cm]{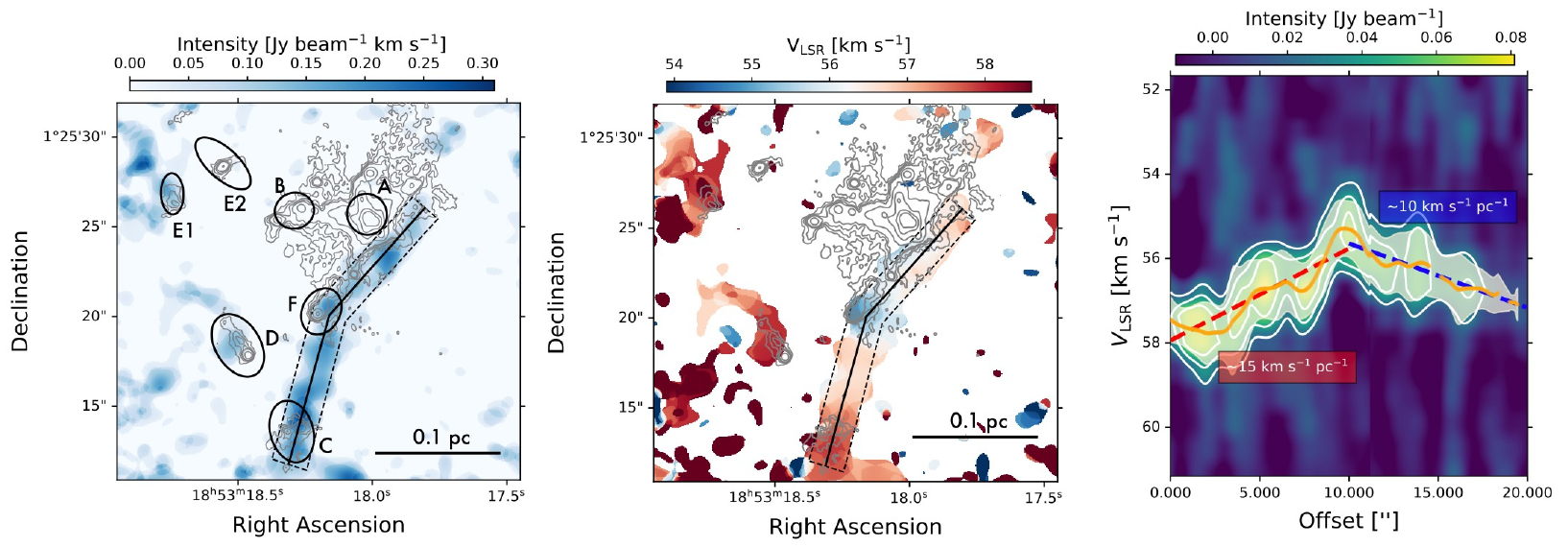}
    \caption{
    The left and middle panels show moment-0 and moment-1 maps of \n2dp, respectively. The overlaid gray contours show Band 6 continuum emission as in Fig.~\ref{fig:continuum}. The letter-labeled black ellipses show the same cores extracted by \citet{2022MNRAS.510.5009L} as in Figure~\ref{fig:continuum}.
    The right panel displays the position-velocity diagram extracted from the region represented by the black solid line and polygon in the left and middle panels. The white contours are at the levels of $\rm [3, 5, 7]\times 8\,\,mJy~beam^{-1}$. 
    The orange solid line is the peak velocity at each offset, and the gray shaded area corresponds to the 1-$\sigma$ velocity dispersion.
    The red and blue dashed lines represent the fitted velocity gradients of the southern and northern filaments, respectively.}
    \label{fig:n2dp}
\end{figure*}

\section{Discussion}
\label{sec:discussion}

\subsection{Stability and Dynamical State}
\label{sec:stability}
Dense cores or condensations will form stars only when they are presently collapsing or will be collapsing in the future. 
To assess the dynamical state of the dense core MM1-C and its two extremely dense substructures (C1 and C2), we estimated their virial parameters. 
We use molecular line data of \n2dp \textit{J}=3-2 from Band 6 observations and \h13cop \textit{J}=1-0 from Band 3 observations \citep{2022MNRAS.510.5009L} to estimate the thermal velocity dispersion $\sigma_{\rm th}$ and non-thermal (turbulence) velocity dispersion $\sigma_{\rm nt}$. The total velocity dispersion is $\sigma_v^2=\sigma_{\rm th}^2+\sigma_{\rm nt}^2$, where $\sigma_{\rm th}=\sqrt{k_B T/\mu m_H}$. On the core scale, we obtain $\sigma_v=0.73\,{\rm km~s}^{-1}$ for \n2dp and $0.46\,{\rm km~s}^{-1}$ for \h13cop. We note that the velocity dispersion measured from gaussian fit of \n2dp should be treated as an upper limit because of its hyperfine structure. We use the same $\sigma_v$ for the two condensations in virial analysis. If we assume the gas temperature is 14.2 K, the thermal sound speed is $c_s=0.22\,{\rm km~s}^{-1}$. The non-thermal velocity dispersion $\sigma_{\rm nt}=0.72\,{\rm km~s}^{-1}$ for \n2dp and 0.45$\,{\rm km~s}^{-1}$ for \h13cop~. The effective sound speed considering thermal and turbulent motion $c_{\rm s}^{\rm eff}=\sqrt{c_s^2+\sigma_{\rm nt}^2}$ is 0.76 and 0.50$\,{\rm km~s}^{-1}$ for \n2dp and \h13cop, respectively. The virial parameter $\alpha_{\rm vir}=M_{\rm vir}/M_{\rm gas}=5\sigma_{\rm 3D}^2 R_{\rm eff}/G M_{\rm gas}$, where the 3D velocity dispersion is $\sigma_{\rm 3D}=\sqrt{3}c_{\rm s}^{\rm eff}$. The derived virial parameters are listed in Table~\ref{tab:dynamics}. We also derive virial parameters assuming a gas temperature of 10$\,$K or 6.5\,K.

The virial parameter of the core MM1-C is $\alpha_{\rm vir}$ = 0.3--0.8, lower than the critical limit, which suggests that the whole core is likely collapsing if additional support from the magnetic field is ignored. 
Therefore, MM1-C is a good candidate to be a massive prestellar core. 
We note that the condensation C1 is trans- or sub-virial except for the temperature of 14.2 K, while C2 is always sub-virial. Note also, that the parent core MM1-C appears more bound than the substructures. This is consistent with \citet{2021MNRAS.502.4963G}, which shows that for any collapsing object with a density profile shallower than $r^{-2}$, the central region will appear unbound. However, half of the super-viral objects in the simulation of \citet{2016ApJ...833..113C} are contracting instead of being unbound, indicating that the external turbulent compressions provide the assembly of the gas. As their masses grow, gravitational contractions will take over.
Since the natal core is in overall collapse, the $M_{\rm gas}$ of C1 and C2 should further grow continuously via gas accretion. Therefore, both C1 and C2 are already contracting or will likely be gravitationally bound in the future. 

A core or condensation will fragment if the physical size is larger than Jeans length, when the internal pressure (thermal+turbulence) can not prevail over gravity. The Jeans length is given as \citep[e.g.,][]{2014MNRAS.439.3275W} 

\begin{align}
    L_{\rm J} &=c^{\rm eff}_s\left(\frac{\pi}{G\rho}\right)^{1/2} \\
    &=0.06\,{\rm pc} \left(\frac{\sigma_v}{0.188\,{\rm km~s}^{-1}}\right)\left(\frac{n}{10^5\, {\rm cm}
^{-3}}\right)^{-1/2},\notag
    \label{equation: jl}
\end{align}
and the corresponding Jeans Mass is given by
\begin{align}
    M_{\rm J}&=\frac{\pi^{5/2}{c_s^{\rm eff}}^3}{6\sqrt{G^3\rho}} \\
    &=0.8\,M_{\odot}\left(\frac{\sigma_v}{0.188\,{\rm km~s}^{-1}}\right)\left(\frac{n}{10^5\, {\rm cm}^{-3}}\right)^{-1/2}. \notag
    \label{equation:mj}
\end{align}

We use \n2dp and \h13cop to determine the stability of C1 and C2. The Jeans lengths and the estimated Jeans masses are summarized in Table~\ref{tab:dynamics}. As the temperature is varied from 6.5 to 14\,K, C1 has a size larger than the Jeans length when $T_{\rm gas}$ is slightly higher than 6.5\,K, suggesting C1 is Jeans unstable only when $T_{\rm gas} \lesssim 6.5$\,K. C2 has a size larger than the Jeans length corresponding to $T_{\rm gas}$ of 10--14\,K, suggesting that C2 is likely Jeans unstable and would undergo further fragmentation.

Interestingly, alongside the two notable condensations C1 and C2, two looming substructures are seen inside C2 with a separation of 0.3$\arcsec$ ($\sim$900 au). As this separation is comparable to the synthesis beam size, \dendro~can not resolve these two peaks. These two substructures indicate that C2 itself is already undergoing further fragmentation, which can be tested by future higher-resolution observations.

Compact substructures have been detected in low-mass prestellar cores such as G205-M3 \citep{2021ApJ...907L..15S}, while fragmentation in massive prestellar cores is rarely reported. We have detected extremely dense substructures within MM1-C, representing the primordial fragmentation within a massive prestellar core. 
Statistic surveys of massive multiple systems reveal that the distribution of the mass ratio $q=M_{\rm comp}/M_{\rm prim}$ exhibits a peak at 0.3--0.5 for wide multiple systems, where $M_{\rm comp}$ is the mass of the companion and $M_{\rm prim}$ is the mass of the primary, accompanied by the negative power-law index\citep[][and references therein]{2013ARA&A..51..269D,2023ASPC..534..275O}. That is to say, the massive wide multiple systems favor low-mass companions.
Therefore we speculate that the massive prestellar core MM1-C will eventually form a multiple stellar system with one massive primary and 1--2 lower-mass companions. However, we cannot fully rule out the possibility that the massive prestellar core MM1-C could undergo further fragmentation and give birth to a cluster of low-mass stars eventually.

\subsection{Cold Filaments Traced by \n2dp }
\label{sec:n2dp}
As shown in Fig.~\ref{fig:spectra}, \n2dp is detected firmly toward core MM1-C. To increase the signal-to-noise ratio, the \n2dp data was smoothed to 1\arcsec~resolution. Shown in Fig.~\ref{fig:n2dp}, MM1-C appears to be situated at one end of a 0.3\,pc-long thin structure traced by \n2dp with a clear velocity gradient. This thin structure seems to trace two filaments connecting dense cores: the southern filament is a gas bridge connecting MM1-C and MM1-F, and the northern filament extends to the filament hub region close to MM1-A.

We extracted a position-velocity (PV) diagram along the filaments. By fitting the PV diagram, we determined the velocity gradients of $\sim$ 15 and 10 $\rm km~s^{-1}~pc^{-1}$ for the southern and northern filaments, respectively. The velocity gradients are quite comparable with the velocity gradients along small-scale filaments in other observations of massive star forming regions~\citep{2021ApJ...915L..10S, 2022MNRAS.514.6038Z, Hacar2023,2023MNRAS.520.3259X} and also in simulations \citep[e.g.][]{2020MNRAS.492.1594S}.
Through the filaments, cold unbounded gas from the envelope of the clump could be channeled to dense cores \citep{Kirk2013}, further increasing their gas masses.

\subsection{Indications to high-mass star formation Models}
Our observations reveal a very promising massive prestellar core (MM1-C) that seems to align with the \textit{Turbulent Core} model. However, the presence of internal fragmentation and the identification of a cold gas filament connecting to the core (MM1-C) challenges the model's expectation of forming a high-mass star through monolithic collapse of an isolated and centrally concentrated core. On the contrary, the presence of a cold filament seem to follow the competitive accretion scenario or other ``clump-fed" models that suggest dynamical mass inflow beyond cores in high-mass star formation. 
However, the initial conditions in those ``clump-feed" models do not favor the existence of a massive prestellar core with such a large gas mass . The complexities in our observations pose questions about the dynamics of massive star formation. The coexistence of massiveness with internal fragmentation and ongoing accretion in core MM1-C underscores the need for more comprehensive comparisons between theories and observations in future studies.

\section{Conclusions}
We present the new ALMA Band 6 continuum and line observations toward the G34-MM1 clump. Accompanied by ATOMS Band 3 data, we report the discovery of two massive prestellar core candidates MM1-C and E1. The line-poor spectra and the absence of outflows suggest their prestellar nature. Their total gas masses are 70 and 20 \solarmass~for MM1-C and E1, respectively. Band 6 observations of $\sim$ 0\farcs3 resolution show that MM1-E1 remains a singular condensation at this resolution. On the other hand, MM1-C is found to be fragmented into two condensations C1 and C2, with mass estimates ranging 1.9--7.0 and 6.5--24.3 \solarmass, respectively, assuming the temperature is in the range of 6.5--14.2\,K. The effective radii are 700 and 1380 au for C1 and C2, resulting in an extremely high density $\rm n_{H_2}\sim 10^{8-9} cm^{-3}$. Condensation C1 is found to be Jeans stable if $T_{\rm gas} \gtrsim 6.5$\,K, while C2 is Jeans unstable when $T_{\rm gas} \lesssim 14.2$\,K.
In addition, C2 seems to have begun to further fragment into two substructures, hinting at its thermal instability. 
We also detected two gas filaments in \n2dp line emission, one connecting core MM1-C to MM1-F, and another extending from MM1-F to the central hub region. Both filaments show a clear velocity gradient of 10--15 $\rm km~s^{-1}~pc^{-1}$. The core MM1-C may further accumulate gas mass along the filament.

To conclude, we detected a fragmenting massive prestellar core candidate, which may give birth to a multiple stellar system containing high-mass star(s).

\section*{Acknowledgements}
This work has been supported by the National Key R$\&$D Program of China (No. 2022YFA1603101), and National Natural Science Foundation of China (NSFC) through grants No. 12073061, 12122307, 12203086, 12103045, 
and 12033005. 

T.L. acknowledges the supports by the international partnership program of Chinese Academy of Sciences through grant No.114231KYSB20200009, and Shanghai Pujiang Program 20PJ1415500.

X.L. is also support by CPSF No. 2022M723278.

H.-L. Liu is supported by Yunnan Fundamental Research Project (grant No.\,202301AT070118).

M.J. acknowledges the support of the Research Council of Finland grant No. 348342.

E.M. is funded by the University of Helsinki doctoral school in particle physics and universe sciences (PAPU).

This research was carried out in part at the Jet Propulsion Laboratory, California Institute of Technology, under a contract with the National Aeronautics and Space Administration (80NM0018D0004).

PS was partially supported by a Grant-in-Aid for Scientific Research (KAKENHI Number JP22H01271 and JP23H01221) of JSPS. 

G.G and L.B. gratefully acknowledge support from the ANID BASAL project FB210003.

B.Z. acknowledges support by the Natural Science Foundation of China (NSFC, grant No. U1831136 and U2031212) and Shanghai Astronomical Observatory (N-2020-06-09-005).

The work of Q-L.G. is supported by Shanghai Rising-Star Program (23YF1455600)

K.T. was supported by JSPS KAKENHI (Grant Number JP20H05645). 

Y.P. Peng acknowledges support from NSFC through grant No. 12303028.

C.W.L. is supported by the Basic Science Research Program through the National Research Foundation of Korea (NRF) funded by the Ministry of Education, Science and Technology (NRF-2019R1A2C1010851) and by the Korea Astronomy and Space Science Institute grant funded by the Korea government (MSIT; Project No. 2023-1-84000).

This paper makes use of the following ALMA data: ADS/JAO.ALMA\#2019.1.00685.S, 2021.1.00095.S and 2023.1.00425. ALMA is a partnership of ESO (representing its member states), NSF (USA) and NINS (Japan), together with NRC (Canada), MOST and ASIAA (Taiwan), and KASI (Republic of Korea), in cooperation with the Republic of Chile. The Joint ALMA Observatory is operated by ESO, AUI/NRAO and NAOJ.

\vspace{5mm}
\facilities{ALMA}

\software{astropy \citep{2013A&A...558A..33A,2018AJ....156..123A},  
          CASA \citep{2022PASP..134k4501C}, 
          CARTA \citep{2021ascl.soft03031C},
          Astrodendro \citep{2008ApJ...679.1338R}}

\appendix
\renewcommand\thefigure{\Alph{section}\arabic{figure}}

\section{Integrated intensity maps of CH$_3$OH lines}
\setcounter{figure}{0}

\begin{figure*}[!thb]
    \centering
    \includegraphics[width=18cm]{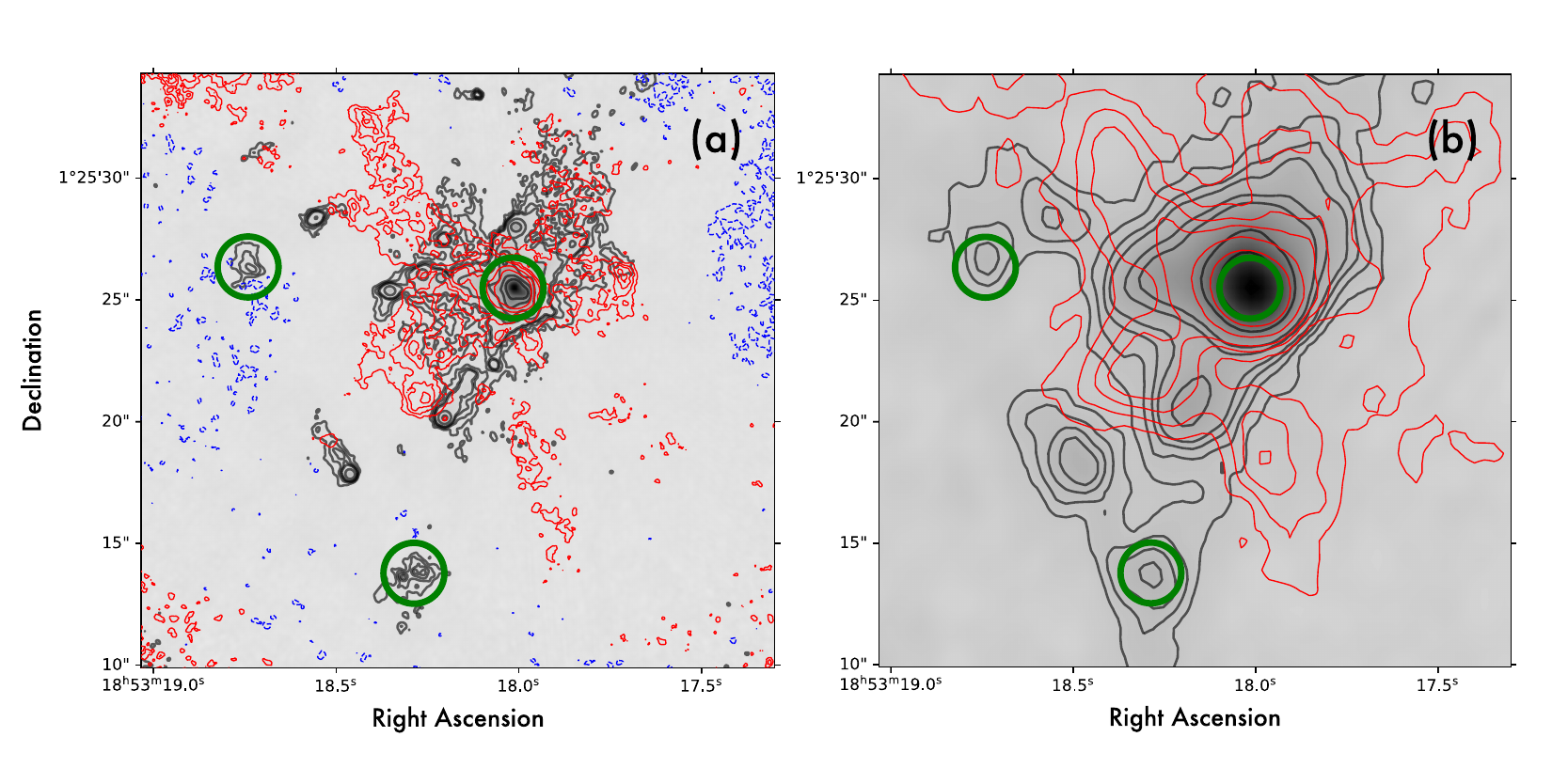}
    \caption{
    Panel (a) shows CH$_3$OH $4_2$-$3_1$ integrated intensity at Band 6. The grayscale background image and black contours are Band 6 continuum emission with the same levels as Fig.~\ref{fig:continuum}. The blue and red contours are at -3$\sigma$ and [3, 5, 7, 9, 15, 30] $\times \sigma$, respectively, where $\rm \sigma=0.08~Jy~beam^{-1}~km~s^{-1}$. Panel (b) shows CH$_3$OH 2(1,1)–1(1,0)A integrate intensity at Band 3. The grayscale background image and black contours are Band 3 continuum emission with the same levels as Fig.~\ref{fig:continuum}. The red contours are at [3, 5, 7, 9, 15, 30] $\times \sigma$, where $\rm \sigma=0.07~Jy~beam^{-1}~km~s^{-1}$. The green circles in both panels represent the areas where the core-averaged spectra in Figure \ref{fig:spectra} were taken.
    }
    \label{fig:appendixa}
\end{figure*}

Fig.~\ref{fig:appendixa} shows the integrated intensity maps of CH$_3$OH $4_2$-$3_1$ (rest frequency: 218.440063 GHz; E$_u$=45.46 K) at Band 6 and another low excitation transition CH$_3$OH 2(1,1)–1(1,0)A (rest frequency: 97.582798 GHz; E$_u$=21.56 K) at Band 3. Panel (a) in Fig.~\ref{fig:appendixa} with negative and positive contours demonstrates that the absorption features like CH$_3$OH seen in the spectra in Fig.~\ref{fig:spectra} could be due to the missing flux issue in interferometric observations. As shown in Panel (b), the low excitation transition line CH$_3$OH 2(1,1)–1(1,0)A as extended as the high excitation one shown in Panel (a), while no emission or absorption is detected toward MM1-C. That means CH$_3$OH is still frozen onto dust grain in such a relatively cold environment. The other molecules such as H$_2$CO and C$^{18}$O are similar to CH$_3$OH toward MM1-C. 

\section{Spectra at the continuum emission peak of MM1-C1}
\label{sec:appb}
\setcounter{figure}{0}

\begin{figure*}[!thb]
    \centering
    \includegraphics[width=18cm]{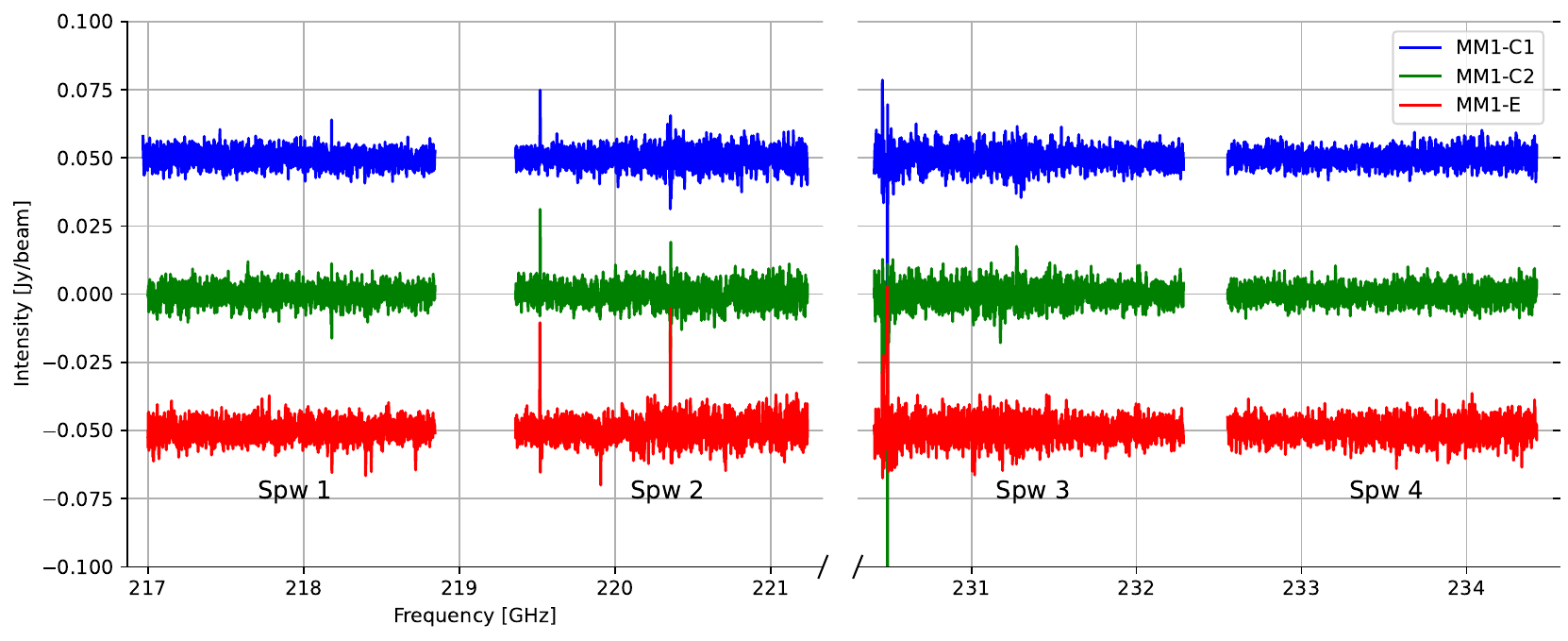}
    \caption{
    Beam-averaged ($0\arcsec.3$ in diameter) spectra at Band 6 centered at the continuum peaks of MM1-C1, MM1-C2 and MM1-E.}
    \label{fig:appendixb}
\end{figure*}

Figure \ref{fig:appendixb} presents the beam-averaged spectra at Band 6 toward the continuum emission peak of MM1-C1. No absorption features are seen in H$_2$CO and CH$_3$OH lines, indicating that there are no warm or hot continuum background sources within MM1-C1. 

\section{CO outflows}
\setcounter{figure}{0} 

To better illustrate the starless characteristics of core MM1-C, we integrated the blue and red wings of the CO high-velocity emission. As Fig.~\ref{fig:appendixb} shows, G34-MM1 clump shows complicated outflow activities, which will be fully explored in a forthcoming paper. However, no CO outflow emission originates from either core MM1-C or MM1-E1.

\begin{figure*}[!thb]
    \centering
    \includegraphics[width=10cm]{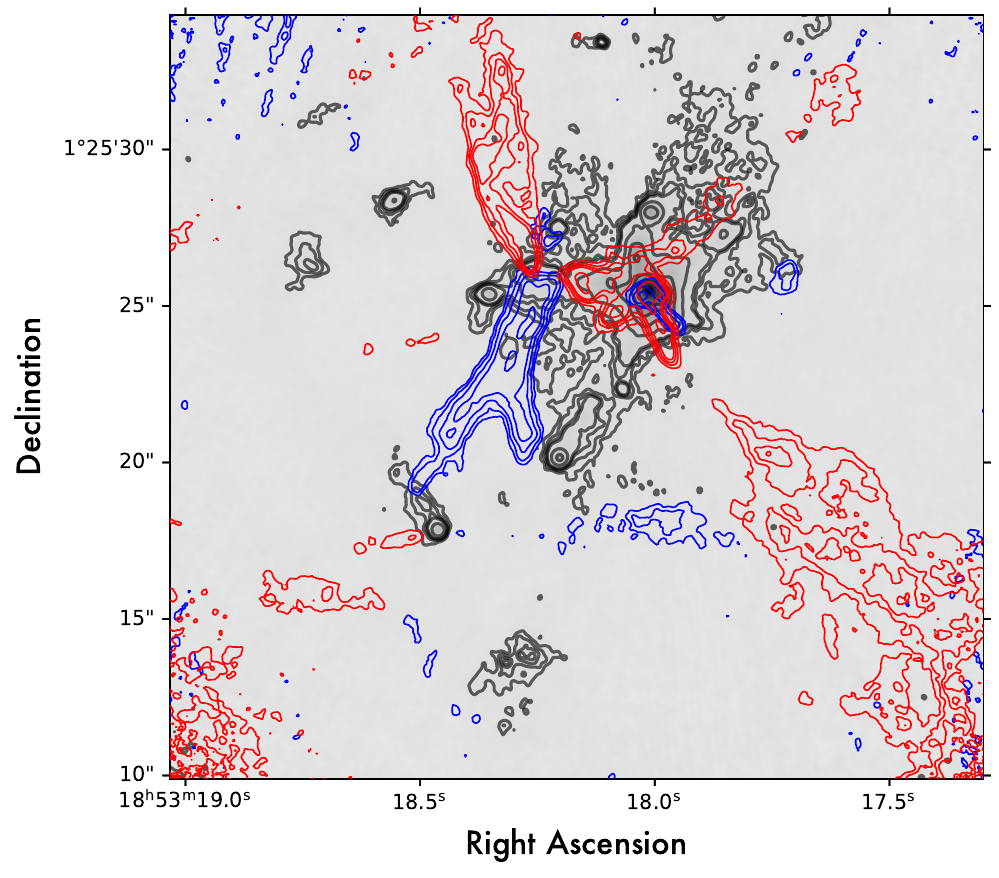}
    \caption{
    Outflow emission traced by CO overlaid on Band 6 continuum emission as in Fig.~\ref{fig:continuum}. Blue and redshifted emission is integrated over [-10, 47] and [67, 107] $\rm km~s^{-1}$. The grayscale background image and black contours are Band 6 continuum emission.
    }
    \label{fig:appendixc}
\end{figure*}

\bibliography{g34starless-v1}{}

\begin{thebibliography}{}
\expandafter\ifx\csname natexlab\endcsname\relax\def\natexlab#1{#1}\fi
\providecommand{\url}[1]{\href{#1}{#1}}
\providecommand{\dodoi}[1]{doi:~\href{http://doi.org/#1}{\nolinkurl{#1}}}
\providecommand{\doeprint}[1]{\href{http://ascl.net/#1}{\nolinkurl{http://ascl.net/#1}}}
\providecommand{\doarXiv}[1]{\href{https://arxiv.org/abs/#1}{\nolinkurl{https://arxiv.org/abs/#1}}}

\bibitem[{{Astropy Collaboration} {et~al.}(2013){Astropy Collaboration},
  {Robitaille}, {Tollerud}, {Greenfield}, {Droettboom}, {Bray}, {Aldcroft},
  {Davis}, {Ginsburg}, {Price-Whelan}, {Kerzendorf}, {Conley}, {Crighton},
  {Barbary}, {Muna}, {Ferguson}, {Grollier}, {Parikh}, {Nair}, {Unther},
  {Deil}, {Woillez}, {Conseil}, {Kramer}, {Turner}, {Singer}, {Fox}, {Weaver},
  {Zabalza}, {Edwards}, {Azalee Bostroem}, {Burke}, {Casey}, {Crawford},
  {Dencheva}, {Ely}, {Jenness}, {Labrie}, {Lim}, {Pierfederici}, {Pontzen},
  {Ptak}, {Refsdal}, {Servillat}, \& {Streicher}}]{2013A&A...558A..33A}
{Astropy Collaboration}, {Robitaille}, T.~P., {Tollerud}, E.~J., {et~al.} 2013,
  \aap, 558, A33, \dodoi{10.1051/0004-6361/201322068}

\bibitem[{{Astropy Collaboration} {et~al.}(2018){Astropy Collaboration},
  {Price-Whelan}, {Sip{\H{o}}cz}, {G{\"u}nther}, {Lim}, {Crawford}, {Conseil},
  {Shupe}, {Craig}, {Dencheva}, {Ginsburg}, {VanderPlas}, {Bradley},
  {P{\'e}rez-Su{\'a}rez}, {de Val-Borro}, {Aldcroft}, {Cruz}, {Robitaille},
  {Tollerud}, {Ardelean}, {Babej}, {Bach}, {Bachetti}, {Bakanov}, {Bamford},
  {Barentsen}, {Barmby}, {Baumbach}, {Berry}, {Biscani}, {Boquien}, {Bostroem},
  {Bouma}, {Brammer}, {Bray}, {Breytenbach}, {Buddelmeijer}, {Burke},
  {Calderone}, {Cano Rodr{\'\i}guez}, {Cara}, {Cardoso}, {Cheedella}, {Copin},
  {Corrales}, {Crichton}, {D'Avella}, {Deil}, {Depagne}, {Dietrich}, {Donath},
  {Droettboom}, {Earl}, {Erben}, {Fabbro}, {Ferreira}, {Finethy}, {Fox},
  {Garrison}, {Gibbons}, {Goldstein}, {Gommers}, {Greco}, {Greenfield},
  {Groener}, {Grollier}, {Hagen}, {Hirst}, {Homeier}, {Horton}, {Hosseinzadeh},
  {Hu}, {Hunkeler}, {Ivezi{\'c}}, {Jain}, {Jenness}, {Kanarek}, {Kendrew},
  {Kern}, {Kerzendorf}, {Khvalko}, {King}, {Kirkby}, {Kulkarni}, {Kumar},
  {Lee}, {Lenz}, {Littlefair}, {Ma}, {Macleod}, {Mastropietro}, {McCully},
  {Montagnac}, {Morris}, {Mueller}, {Mumford}, {Muna}, {Murphy}, {Nelson},
  {Nguyen}, {Ninan}, {N{\"o}the}, {Ogaz}, {Oh}, {Parejko}, {Parley}, {Pascual},
  {Patil}, {Patil}, {Plunkett}, {Prochaska}, {Rastogi}, {Reddy Janga},
  {Sabater}, {Sakurikar}, {Seifert}, {Sherbert}, {Sherwood-Taylor}, {Shih},
  {Sick}, {Silbiger}, {Singanamalla}, {Singer}, {Sladen}, {Sooley},
  {Sornarajah}, {Streicher}, {Teuben}, {Thomas}, {Tremblay}, {Turner},
  {Terr{\'o}n}, {van Kerkwijk}, {de la Vega}, {Watkins}, {Weaver}, {Whitmore},
  {Woillez}, {Zabalza}, \& {Astropy Contributors}}]{2018AJ....156..123A}
{Astropy Collaboration}, {Price-Whelan}, A.~M., {Sip{\H{o}}cz}, B.~M., {et~al.}
  2018, \aj, 156, 123, \dodoi{10.3847/1538-3881/aabc4f}

\bibitem[{{Barnes} {et~al.}(2023){Barnes}, {Liu}, {Zhang}, {Tan}, {Bigiel},
  {Caselli}, {Cosentino}, {Fontani}, {Henshaw}, {Jim{\'e}nez-Serra}, {Kalb},
  {Law}, {Longmore}, {Parker}, {Pineda}, {S{\'a}nchez-Monge}, {Lim}, \&
  {Wang}}]{2023A&A...675A..53B}
{Barnes}, A.~T., {Liu}, J., {Zhang}, Q., {et~al.} 2023, \aap, 675, A53,
  \dodoi{10.1051/0004-6361/202245668}

\bibitem[{{Benedettini} {et~al.}(2018){Benedettini}, {Pezzuto}, {Schisano},
  {Andr{\'e}}, {K{\"o}nyves}, {Men'shchikov}, {Ladjelate}, {Di Francesco},
  {Elia}, {Arzoumanian}, {Louvet}, {Palmeirim}, {Rygl}, {Schneider},
  {Spinoglio}, \& {Ward-Thompson}}]{Benedettini2018}
{Benedettini}, M., {Pezzuto}, S., {Schisano}, E., {et~al.} 2018, \aap, 619,
  A52, \dodoi{10.1051/0004-6361/201833364}

\bibitem[{{Bonnell} {et~al.}(2001){Bonnell}, {Bate}, {Clarke}, \&
  {Pringle}}]{2001MNRAS.323..785B}
{Bonnell}, I.~A., {Bate}, M.~R., {Clarke}, C.~J., \& {Pringle}, J.~E. 2001,
  \mnras, 323, 785, \dodoi{10.1046/j.1365-8711.2001.04270.x}

\bibitem[{{Bontemps} {et~al.}(2010){Bontemps}, {Motte}, {Csengeri}, \&
  {Schneider}}]{2010A&A...524A..18B}
{Bontemps}, S., {Motte}, F., {Csengeri}, T., \& {Schneider}, N. 2010, \aap,
  524, A18, \dodoi{10.1051/0004-6361/200913286}

\bibitem[{{Camacho} {et~al.}(2016){Camacho}, {V{\'a}zquez-Semadeni},
  {Ballesteros-Paredes}, {G{\'o}mez}, {Fall}, \&
  {Mata-Ch{\'a}vez}}]{2016ApJ...833..113C}
{Camacho}, V., {V{\'a}zquez-Semadeni}, E., {Ballesteros-Paredes}, J., {et~al.}
  2016, \apj, 833, 113, \dodoi{10.3847/1538-4357/833/1/113}

\bibitem[{{CASA Team} {et~al.}(2022){CASA Team}, {Bean}, {Bhatnagar}, {Castro},
  {Donovan Meyer}, {Emonts}, {Garcia}, {Garwood}, {Golap}, {Gonzalez Villalba},
  {Harris}, {Hayashi}, {Hoskins}, {Hsieh}, {Jagannathan}, {Kawasaki},
  {Keimpema}, {Kettenis}, {Lopez}, {Marvil}, {Masters}, {McNichols},
  {Mehringer}, {Miel}, {Moellenbrock}, {Montesino}, {Nakazato}, {Ott}, {Petry},
  {Pokorny}, {Raba}, {Rau}, {Schiebel}, {Schweighart}, {Sekhar}, {Shimada},
  {Small}, {Steeb}, {Sugimoto}, {Suoranta}, {Tsutsumi}, {van Bemmel},
  {Verkouter}, {Wells}, {Xiong}, {Szomoru}, {Griffith}, {Glendenning}, \&
  {Kern}}]{2022PASP..134k4501C}
{CASA Team}, {Bean}, B., {Bhatnagar}, S., {et~al.} 2022, \pasp, 134, 114501,
  \dodoi{10.1088/1538-3873/ac9642}

\bibitem[{{Caselli} {et~al.}(2019){Caselli}, {Pineda}, {Zhao}, {Walmsley},
  {Keto}, {Tafalla}, {Chac{\'o}n-Tanarro}, {Bourke}, {Friesen}, {Galli}, \&
  {Padovani}}]{2019ApJ...874...89C}
{Caselli}, P., {Pineda}, J.~E., {Zhao}, B., {et~al.} 2019, \apj, 874, 89,
  \dodoi{10.3847/1538-4357/ab0700}

\bibitem[{{Comrie} {et~al.}(2021){Comrie}, {Wang}, {Hsu}, {Moraghan}, {Harris},
  {Pang}, {Pi{\'n}ska}, {Chiang}, {Simmonds}, {Chang}, {Jan}, \&
  {Lin}}]{2021ascl.soft03031C}
{Comrie}, A., {Wang}, K.-S., {Hsu}, S.-C., {et~al.} 2021, {CARTA: Cube Analysis
  and Rendering Tool for Astronomy}, Astrophysics Source Code Library, record
  ascl:2103.031.
\newblock \doeprint{2103.031}

\bibitem[{{Crapsi} {et~al.}(2007){Crapsi}, {Caselli}, {Walmsley}, \&
  {Tafalla}}]{2007A&A...470..221C}
{Crapsi}, A., {Caselli}, P., {Walmsley}, M.~C., \& {Tafalla}, M. 2007, \aap,
  470, 221, \dodoi{10.1051/0004-6361:20077613}

\bibitem[{{Duch{\^e}ne} \& {Kraus}(2013)}]{2013ARA&A..51..269D}
{Duch{\^e}ne}, G., \& {Kraus}, A. 2013, \araa, 51, 269,
  \dodoi{10.1146/annurev-astro-081710-102602}

\bibitem[{{Dzib} {et~al.}(2023){Dzib}, {Yang}, {Urquhart}, {Medina},
  {Brunthaler}, {Menten}, {Wyrowski}, {Cotton}, {Dokara}, {Ortiz-Le{\'o}n},
  {Rugel}, {Nguyen}, {Gong}, {Chakraborty}, {Beuther}, {Billington},
  {Carrasco-Gonzalez}, {Csengeri}, {Hofner}, {Ott}, {Pandian}, {Roy}, \&
  {Yanza}}]{2023A&A...670A...9D}
{Dzib}, S.~A., {Yang}, A.~Y., {Urquhart}, J.~S., {et~al.} 2023, \aap, 670, A9,
  \dodoi{10.1051/0004-6361/202143019}

\bibitem[{{Foster} {et~al.}(2014){Foster}, {Arce}, {Kassis}, {Sanhueza},
  {Jackson}, {Finn}, {Offner}, {Sakai}, {Sakai}, {Yamamoto}, {Guzm{\'a}n}, \&
  {Rathborne}}]{2014ApJ...791..108F}
{Foster}, J.~B., {Arce}, H.~G., {Kassis}, M., {et~al.} 2014, \apj, 791, 108,
  \dodoi{10.1088/0004-637X/791/2/108}

\bibitem[{{Garay} {et~al.}(2004){Garay}, {Fa{\'u}ndez}, {Mardones}, {Bronfman},
  {Chini}, \& {Nyman}}]{2004ApJ...610..313G}
{Garay}, G., {Fa{\'u}ndez}, S., {Mardones}, D., {et~al.} 2004, \apj, 610, 313,
  \dodoi{10.1086/421437}

\bibitem[{{G{\'o}mez} {et~al.}(2021){G{\'o}mez}, {V{\'a}zquez-Semadeni}, \&
  {Palau}}]{2021MNRAS.502.4963G}
{G{\'o}mez}, G.~C., {V{\'a}zquez-Semadeni}, E., \& {Palau}, A. 2021, \mnras,
  502, 4963, \dodoi{10.1093/mnras/stab394}

\bibitem[{{Hacar} {et~al.}(2023){Hacar}, {Clark}, {Heitsch}, {Kainulainen},
  {Panopoulou}, {Seifried}, \& {Smith}}]{Hacar2023}
{Hacar}, A., {Clark}, S.~E., {Heitsch}, F., {et~al.} 2023, in Astronomical
  Society of the Pacific Conference Series, Vol. 534, Protostars and Planets
  VII, ed. S.~{Inutsuka}, Y.~{Aikawa}, T.~{Muto}, K.~{Tomida}, \& M.~{Tamura},
  153, \dodoi{10.48550/arXiv.2203.09562}

\bibitem[{{Kirk} {et~al.}(2013){Kirk}, {Myers}, {Bourke}, {Gutermuth},
  {Hedden}, \& {Wilson}}]{Kirk2013}
{Kirk}, H., {Myers}, P.~C., {Bourke}, T.~L., {et~al.} 2013, \apj, 766, 115,
  \dodoi{10.1088/0004-637X/766/2/115}

\bibitem[{{Li} {et~al.}(2023){Li}, {Sanhueza}, {Zhang}, {Guido}, {Sabatini},
  {Morii}, {Lu}, {Tafoya}, {Nakamura}, {Izumi}, {Tatematsu}, \&
  {Li}}]{2023ApJ...949..109L}
{Li}, S., {Sanhueza}, P., {Zhang}, Q., {et~al.} 2023, \apj, 949, 109,
  \dodoi{10.3847/1538-4357/acc58f}

\bibitem[{{Liu} {et~al.}(2020{\natexlab{a}}){Liu}, {Sanhueza}, {Liu},
  {Zavagno}, {Tang}, {Wu}, \& {Zhang}}]{2020ApJ...901...31L}
{Liu}, H.-L., {Sanhueza}, P., {Liu}, T., {et~al.} 2020{\natexlab{a}}, \apj,
  901, 31, \dodoi{10.3847/1538-4357/abadfe}

\bibitem[{{Liu} {et~al.}(2021{\natexlab{a}}){Liu}, {Liu}, {Evans}, {Wang},
  {Garay}, {Qin}, {Li}, {Stutz}, {Goldsmith}, {Liu}, {Tej}, {Zhang}, {Juvela},
  {Li}, {Wang}, {Bronfman}, {Ren}, {Wu}, {Kim}, {Lee}, {Tatematsu},
  {Cunningham}, {Liu}, {Wu}, {Hirota}, {Lee}, {Li}, {Kang}, {Mardones},
  {Ristorcelli}, {Zhang}, {Luo}, {Toth}, {Yi}, {Yun}, {Peng}, {Li}, {Zhu},
  {Shen}, {Baug}, {Dewangan}, {Chakali}, {Liu}, {Xu}, {Wang}, {Zhang}, {Li},
  {Zhang}, {Zhou}, {Tang}, {Xue}, {Issac}, {Soam}, \&
  {{\'A}lvarez-Guti{\'e}rrez}}]{2021MNRAS.505.2801L}
{Liu}, H.-L., {Liu}, T., {Evans}, Neal~J., I., {et~al.} 2021{\natexlab{a}},
  \mnras, 505, 2801, \dodoi{10.1093/mnras/stab1352}

\bibitem[{{Liu} {et~al.}(2022){Liu}, {Tej}, {Liu}, {Issac}, {Saha},
  {Goldsmith}, {Wang}, {Zhang}, {Qin}, {Wang}, {Li}, {Soam}, {Dewangan}, {Lee},
  {Li}, {Liu}, {Zhang}, {Ren}, {Juvela}, {Bronfman}, {Wu}, {Tatematsu}, {Chen},
  {Li}, {Stutz}, {Zhang}, {Viktor Toth}, {Luo}, {Xu}, {Li}, {Liu}, {Zhou},
  {Zhang}, {Tang}, {Zhang}, {Baug}, {Mannfors}, {Chakali}, \&
  {Dutta}}]{2022MNRAS.510.5009L}
{Liu}, H.-L., {Tej}, A., {Liu}, T., {et~al.} 2022, \mnras, 510, 5009,
  \dodoi{10.1093/mnras/stab2757}

\bibitem[{{Liu} {et~al.}(2017){Liu}, {Lacy}, {Li}, {Wang}, {Qin}, {Zhang},
  {Kim}, {Garay}, {Wu}, {Mardones}, {Zhu}, {Tatematsu}, {Hirota}, {Ren}, {Liu},
  {Chen}, {Su}, \& {Li}}]{Liu2017}
{Liu}, T., {Lacy}, J., {Li}, P.~S., {et~al.} 2017, \apj, 849, 25,
  \dodoi{10.3847/1538-4357/aa8d73}

\bibitem[{{Liu} {et~al.}(2020{\natexlab{b}}){Liu}, {Evans}, {Kim}, {Goldsmith},
  {Liu}, {Zhang}, {Tatematsu}, {Wang}, {Juvela}, {Bronfman}, {Cunningham},
  {Garay}, {Hirota}, {Lee}, {Kang}, {Li}, {Li}, {Mardones}, {Qin},
  {Ristorcelli}, {Tej}, {Toth}, {Wu}, {Wu}, {Yi}, {Yun}, {Liu}, {Peng}, {Li},
  {Li}, {Lee}, {Shen}, {Baug}, {Wang}, {Zhang}, {Issac}, {Zhu}, {Luo}, {Soam},
  {Liu}, {Xu}, {Wang}, {Zhang}, {Ren}, \& {Zhang}}]{2020MNRAS.496.2790L}
{Liu}, T., {Evans}, N.~J., {Kim}, K.-T., {et~al.} 2020{\natexlab{b}}, \mnras,
  496, 2790, \dodoi{10.1093/mnras/staa1577}

\bibitem[{Liu {et~al.}(2023)Liu, Liu, Zhu, Garay, Liu, Goldsmith, Evans, Kim,
  Liu, Xu, Lu, Tej, Mai, Bronfman, Li, Mardones, Stutz, Tatematsu, Wang, Zhang,
  Qin, Zhou, Luo, Zhang, Cheng, He, Gu, Li, Zhang, Zhang, Saha, Dewangan,
  Sanhueza, \& Shen}]{liu2023almaquarks}
Liu, X., Liu, T., Zhu, L., {et~al.} 2023, The ALMA-QUARKS survey: -- I. Survey
  description and data reduction.
\newblock \doarXiv{2311.08651}

\bibitem[{{Liu} {et~al.}(2021{\natexlab{b}}){Liu}, {Wu}, {Zhang}, {Chen},
  {Lin}, {Qin}, {Liu}, {Henkel}, {Wang}, {Liu}, {Yuan}, {Yuan}, {Li}, {Shen},
  {Li}, {Esimbek}, {Wang}, {Li}, {Kim}, {Zhu}, {Madones}, {Inostroza-Pino},
  {Meng}, {Zhang}, {Tatematsu}, {Xu}, {Ju}, {Kraus}, \&
  {Xu}}]{2021ApJ...912..148L}
{Liu}, X.~C., {Wu}, Y., {Zhang}, C., {et~al.} 2021{\natexlab{b}}, \apj, 912,
  148, \dodoi{10.3847/1538-4357/abee73}

\bibitem[{{Lu} {et~al.}(2014){Lu}, {Zhang}, {Liu}, {Wang}, \&
  {Gu}}]{2014ApJ...790...84L}
{Lu}, X., {Zhang}, Q., {Liu}, H.~B., {Wang}, J., \& {Gu}, Q. 2014, \apj, 790,
  84, \dodoi{10.1088/0004-637X/790/2/84}

\bibitem[{{Mai} {et~al.}(2023){Mai}, {Zhang}, {Reid}, {Moscadelli}, {Xu},
  {Sun}, {Zhang}, {Chen}, {Wen}, {Luo}, {Menten}, {Zheng}, {Brunthaler}, {Xu},
  \& {Wang}}]{2023ApJ...949...10M}
{Mai}, X., {Zhang}, B., {Reid}, M.~J., {et~al.} 2023, \apj, 949, 10,
  \dodoi{10.3847/1538-4357/acc52a}

\bibitem[{{McKee} \& {Tan}(2003)}]{2003ApJ...585..850M}
{McKee}, C.~F., \& {Tan}, J.~C. 2003, \apj, 585, 850, \dodoi{10.1086/346149}

\bibitem[{{Morii} {et~al.}(2021){Morii}, {Sanhueza}, {Nakamura}, {Jackson},
  {Li}, {Beuther}, {Zhang}, {Feng}, {Tafoya}, {Guzm{\'a}n}, {Izumi}, {Sakai},
  {Lu}, {Tatematsu}, {Ohashi}, {Silva}, {Olguin}, \&
  {Contreras}}]{2021ApJ...923..147M}
{Morii}, K., {Sanhueza}, P., {Nakamura}, F., {et~al.} 2021, \apj, 923, 147,
  \dodoi{10.3847/1538-4357/ac2365}

\bibitem[{{Morii} {et~al.}(2023){Morii}, {Sanhueza}, {Nakamura}, {Zhang},
  {Sabatini}, {Beuther}, {Lu}, {Li}, {Garay}, {Jackson}, {Olguin}, {Tafoya},
  {Tatematsu}, {Izumi}, {Sakai}, \& {Silva}}]{2023ApJ...950..148M}
---. 2023, \apj, 950, 148, \dodoi{10.3847/1538-4357/acccea}

\bibitem[{{Motte} {et~al.}(2018){Motte}, {Nony}, {Louvet}, {Marsh}, {Bontemps},
  {Whitworth}, {Men'shchikov}, {Nguyen Luong}, {Csengeri}, {Maury}, {Gusdorf},
  {Chapillon}, {K{\"o}nyves}, {Schilke}, {Duarte-Cabral}, {Didelon}, \&
  {Gaudel}}]{Motte2018}
{Motte}, F., {Nony}, T., {Louvet}, F., {et~al.} 2018, Nature Astronomy, 2, 478,
  \dodoi{10.1038/s41550-018-0452-x}

\bibitem[{{Nony} {et~al.}(2018){Nony}, {Louvet}, {Motte}, {Molet}, {Marsh},
  {Chapillon}, {Gusdorf}, {Brouillet}, {Bontemps}, {Csengeri}, {Despois},
  {Nguyen Luong}, {Duarte-Cabral}, \& {Maury}}]{2018A&A...618L...5N}
{Nony}, T., {Louvet}, F., {Motte}, F., {et~al.} 2018, \aap, 618, L5,
  \dodoi{10.1051/0004-6361/201833863}

\bibitem[{{Offner} {et~al.}(2023){Offner}, {Moe}, {Kratter}, {Sadavoy},
  {Jensen}, \& {Tobin}}]{2023ASPC..534..275O}
{Offner}, S.~S.~R., {Moe}, M., {Kratter}, K.~M., {et~al.} 2023, in Astronomical
  Society of the Pacific Conference Series, Vol. 534, Protostars and Planets
  VII, ed. S.~{Inutsuka}, Y.~{Aikawa}, T.~{Muto}, K.~{Tomida}, \& M.~{Tamura},
  275, \dodoi{10.48550/arXiv.2203.10066}

\bibitem[{{Ossenkopf} \& {Henning}(1994)}]{1994A&A...291..943O}
{Ossenkopf}, V., \& {Henning}, T. 1994, \aap, 291, 943

\bibitem[{{Padoan} {et~al.}(2020){Padoan}, {Pan}, {Juvela}, {Haugb{\o}lle}, \&
  {Nordlund}}]{2020ApJ...900...82P}
{Padoan}, P., {Pan}, L., {Juvela}, M., {Haugb{\o}lle}, T., \& {Nordlund},
  {\r{A}}. 2020, \apj, 900, 82, \dodoi{10.3847/1538-4357/abaa47}

\bibitem[{{Palau} {et~al.}(2013){Palau}, {Fuente}, {Girart}, {Estalella}, {Ho},
  {S{\'a}nchez-Monge}, {Fontani}, {Busquet}, {Commer{\c{c}}on}, {Hennebelle},
  {Boissier}, {Zhang}, {Cesaroni}, \& {Zapata}}]{2013ApJ...762..120P}
{Palau}, A., {Fuente}, A., {Girart}, J.~M., {et~al.} 2013, \apj, 762, 120,
  \dodoi{10.1088/0004-637X/762/2/120}

\bibitem[{{Pollack} {et~al.}(1994){Pollack}, {Hollenbach}, {Beckwith},
  {Simonelli}, {Roush}, \& {Fong}}]{1994ApJ...421..615P}
{Pollack}, J.~B., {Hollenbach}, D., {Beckwith}, S., {et~al.} 1994, \apj, 421,
  615, \dodoi{10.1086/173677}

\bibitem[{{Pouteau} {et~al.}(2022){Pouteau}, {Motte}, {Nony},
  {Galv{\'a}n-Madrid}, {Men'shchikov}, {Bontemps}, {Robitaille}, {Louvet},
  {Ginsburg}, {Herpin}, {L{\'o}pez-Sepulcre}, {Dell'Ova}, {Gusdorf},
  {Sanhueza}, {Stutz}, {Brouillet}, {Thomasson}, {Armante}, {Baug}, {Bonfand},
  {Busquet}, {Csengeri}, {Cunningham}, {Fern{\'a}ndez-L{\'o}pez}, {Liu},
  {Olguin}, {Towner}, {Bally}, {Braine}, {Bronfman}, {Joncour}, {Gonz{\'a}lez},
  {Hennebelle}, {Lu}, {Menten}, {Moraux}, {Tatematsu}, {Walker}, \&
  {Whitworth}}]{Pouteau2022}
{Pouteau}, Y., {Motte}, F., {Nony}, T., {et~al.} 2022, \aap, 664, A26,
  \dodoi{10.1051/0004-6361/202142951}

\bibitem[{{Qin} {et~al.}(2022){Qin}, {Liu}, {Liu}, {Goldsmith}, {Li}, {Zhang},
  {Liu}, {Wu}, {Bronfman}, {Juvela}, {Lee}, {Garay}, {Zhang}, {He}, {Hsu},
  {Shen}, {Lee}, {Wang}, {Tang}, {Tang}, {Zhang}, {Yue}, {Xue}, {Li}, {Peng},
  {Dutta}, {Ge}, {Xu}, {Chen}, {Baug}, {Dewangan}, \&
  {Tej}}]{2022MNRAS.511.3463Q}
{Qin}, S.-L., {Liu}, T., {Liu}, X., {et~al.} 2022, \mnras, 511, 3463,
  \dodoi{10.1093/mnras/stac219}

\bibitem[{{Rathborne} {et~al.}(2005){Rathborne}, {Jackson}, {Chambers},
  {Simon}, {Shipman}, \& {Frieswijk}}]{2005ApJ...630L.181R}
{Rathborne}, J.~M., {Jackson}, J.~M., {Chambers}, E.~T., {et~al.} 2005, \apjl,
  630, L181, \dodoi{10.1086/491656}

\bibitem[{{Rosolowsky} {et~al.}(2008){Rosolowsky}, {Pineda}, {Kauffmann}, \&
  {Goodman}}]{2008ApJ...679.1338R}
{Rosolowsky}, E.~W., {Pineda}, J.~E., {Kauffmann}, J., \& {Goodman}, A.~A.
  2008, \apj, 679, 1338, \dodoi{10.1086/587685}

\bibitem[{{Sahu} {et~al.}(2021){Sahu}, {Liu}, {Liu}, {Evans}, {Hirano},
  {Tatematsu}, {Lee}, {Kim}, {Dutta}, {Alina}, {Bronfman}, {Cunningham},
  {Eden}, {Garay}, {Goldsmith}, {He}, {Hsu}, {Jhan}, {Johnstone}, {Juvela},
  {Kim}, {Kuan}, {Kwon}, {Lee}, {Lee}, {Li}, {Li}, {Li}, {Luo}, {Montillaud},
  {Moraghan}, {Pelkonen}, {Qin}, {Ristorcelli}, {Sanhueza}, {Shang}, {Shen},
  {Soam}, {Wu}, {Zhang}, \& {Zhou}}]{2021ApJ...907L..15S}
{Sahu}, D., {Liu}, S.-Y., {Liu}, T., {et~al.} 2021, \apjl, 907, L15,
  \dodoi{10.3847/2041-8213/abd3aa}

\bibitem[{{Sakai} {et~al.}(2013){Sakai}, {Sakai}, {Foster}, {Sanhueza},
  {Jackson}, {Kassis}, {Furuya}, {Aikawa}, {Hirota}, \&
  {Yamamoto}}]{2013ApJ...775L..31S}
{Sakai}, T., {Sakai}, N., {Foster}, J.~B., {et~al.} 2013, \apjl, 775, L31,
  \dodoi{10.1088/2041-8205/775/1/L31}

\bibitem[{{Sanhueza} {et~al.}(2010){Sanhueza}, {Garay}, {Bronfman}, {Mardones},
  {May}, \& {Saito}}]{2010ApJ...715...18S}
{Sanhueza}, P., {Garay}, G., {Bronfman}, L., {et~al.} 2010, \apj, 715, 18,
  \dodoi{10.1088/0004-637X/715/1/18}

\bibitem[{{Sanhueza} {et~al.}(2012){Sanhueza}, {Jackson}, {Foster}, {Garay},
  {Silva}, \& {Finn}}]{2012ApJ...756...60S}
{Sanhueza}, P., {Jackson}, J.~M., {Foster}, J.~B., {et~al.} 2012, \apj, 756,
  60, \dodoi{10.1088/0004-637X/756/1/60}

\bibitem[{{Sanhueza} {et~al.}(2017){Sanhueza}, {Jackson}, {Zhang},
  {Guzm{\'a}n}, {Lu}, {Stephens}, {Wang}, \& {Tatematsu}}]{2017ApJ...841...97S}
{Sanhueza}, P., {Jackson}, J.~M., {Zhang}, Q., {et~al.} 2017, \apj, 841, 97,
  \dodoi{10.3847/1538-4357/aa6ff8}

\bibitem[{{Sanhueza} {et~al.}(2019){Sanhueza}, {Contreras}, {Wu}, {Jackson},
  {Guzm{\'a}n}, {Zhang}, {Li}, {Lu}, {Silva}, {Izumi}, {Liu}, {Miura},
  {Tatematsu}, {Sakai}, {Beuther}, {Garay}, {Ohashi}, {Saito}, {Nakamura},
  {Saigo}, {Veena}, {Nguyen-Luong}, \& {Tafoya}}]{2019ApJ...886..102S}
{Sanhueza}, P., {Contreras}, Y., {Wu}, B., {et~al.} 2019, \apj, 886, 102,
  \dodoi{10.3847/1538-4357/ab45e9}

\bibitem[{{Sanhueza} {et~al.}(2021){Sanhueza}, {Girart}, {Padovani}, {Galli},
  {Hull}, {Zhang}, {Cortes}, {Stephens}, {Fern{\'a}ndez-L{\'o}pez}, {Jackson},
  {Frau}, {Kock}, {Wu}, {Zapata}, {Olguin}, {Lu}, {Silva}, {Tang}, {Sakai},
  {Guzm{\'a}n}, {Tatematsu}, {Nakamura}, \& {Chen}}]{2021ApJ...915L..10S}
{Sanhueza}, P., {Girart}, J.~M., {Padovani}, M., {et~al.} 2021, \apjl, 915,
  L10, \dodoi{10.3847/2041-8213/ac081c}

\bibitem[{{Simon} {et~al.}(2001){Simon}, {Jackson}, {Clemens}, {Bania}, \&
  {Heyer}}]{2001ApJ...551..747S}
{Simon}, R., {Jackson}, J.~M., {Clemens}, D.~P., {Bania}, T.~M., \& {Heyer},
  M.~H. 2001, \apj, 551, 747, \dodoi{10.1086/320230}

\bibitem[{{Simon} {et~al.}(2006){Simon}, {Rathborne}, {Shah}, {Jackson}, \&
  {Chambers}}]{2006ApJ...653.1325S}
{Simon}, R., {Rathborne}, J.~M., {Shah}, R.~Y., {Jackson}, J.~M., \&
  {Chambers}, E.~T. 2006, \apj, 653, 1325, \dodoi{10.1086/508915}

\bibitem[{{Smith} {et~al.}(2020){Smith}, {Tre{\ss}}, {Sormani}, {Glover},
  {Klessen}, {Clark}, {Izquierdo}, {Duarte-Cabral}, \&
  {Zucker}}]{2020MNRAS.492.1594S}
{Smith}, R.~J., {Tre{\ss}}, R.~G., {Sormani}, M.~C., {et~al.} 2020, \mnras,
  492, 1594, \dodoi{10.1093/mnras/stz3328}

\bibitem[{{Soam} {et~al.}(2019){Soam}, {Liu}, {Andersson}, {Lee}, {Liu},
  {Juvela}, {Li}, {Goldsmith}, {Zhang}, {Koch}, {Kim}, {Qiu}, {Evans},
  {Johnstone}, {Thompson}, {Ward-Thompson}, {Di Francesco}, {Tang},
  {Montillaud}, {Kim}, {Mairs}, {Sanhueza}, {Kim}, {Berry}, {Gordon},
  {Tatematsu}, {Liu}, {Pattle}, {Eden}, {McGehee}, {Wang}, {Ristorcelli},
  {Graves}, {Alina}, {Lacaille}, {Montier}, {Park}, {Kwon}, {Chung},
  {Pelkonen}, {Micelotta}, {Saajasto}, \& {Fuller}}]{2019ApJ...883...95S}
{Soam}, A., {Liu}, T., {Andersson}, B.~G., {et~al.} 2019, \apj, 883, 95,
  \dodoi{10.3847/1538-4357/ab39dd}

\bibitem[{{Spear} {et~al.}(2021){Spear}, {Maureira}, {Arce}, {Pineda},
  {Dunham}, {Caselli}, \& {Segura-Cox}}]{2021ApJ...923..231S}
{Spear}, S., {Maureira}, M.~J., {Arce}, H.~G., {et~al.} 2021, \apj, 923, 231,
  \dodoi{10.3847/1538-4357/ac3083}

\bibitem[{{V{\'a}zquez-Semadeni} {et~al.}(2009){V{\'a}zquez-Semadeni},
  {G{\'o}mez}, {Jappsen}, {Ballesteros-Paredes}, \&
  {Klessen}}]{2009ApJ...707.1023V}
{V{\'a}zquez-Semadeni}, E., {G{\'o}mez}, G.~C., {Jappsen}, A.~K.,
  {Ballesteros-Paredes}, J., \& {Klessen}, R.~S. 2009, \apj, 707, 1023,
  \dodoi{10.1088/0004-637X/707/2/1023}

\bibitem[{{V{\'a}zquez-Semadeni} {et~al.}(2019){V{\'a}zquez-Semadeni}, {Palau},
  {Ballesteros-Paredes}, {G{\'o}mez}, \&
  {Zamora-Avil{\'e}s}}]{2019MNRAS.490.3061V}
{V{\'a}zquez-Semadeni}, E., {Palau}, A., {Ballesteros-Paredes}, J.,
  {G{\'o}mez}, G.~C., \& {Zamora-Avil{\'e}s}, M. 2019, \mnras, 490, 3061,
  \dodoi{10.1093/mnras/stz2736}

\bibitem[{{Wang} {et~al.}(2014){Wang}, {Zhang}, {Testi}, {van der Tak}, {Wu},
  {Zhang}, {Pillai}, {Wyrowski}, {Carey}, {Ragan}, \&
  {Henning}}]{2014MNRAS.439.3275W}
{Wang}, K., {Zhang}, Q., {Testi}, L., {et~al.} 2014, \mnras, 439, 3275,
  \dodoi{10.1093/mnras/stu127}

\bibitem[{{Xu} {et~al.}(2023{\natexlab{a}}){Xu}, {Wang}, {Liu}, {Tang},
  {Evans}, {Palau}, {Morii}, {He}, {Sanhueza}, {Liu}, {Stutz}, {Zhang}, {Chen},
  {Li}, {G{\'o}mez}, {V{\'a}zquez-Semadeni}, {Li}, {Mai}, {Lu}, {Liu}, {Chen},
  {Li}, {Shi}, {Ren}, {Li}, {Garay}, {Bronfman}, {Dewangan}, {Juvela}, {Lee},
  {Zhang}, {Yue}, {Wang}, {Ge}, {Jiao}, {Luo}, {Zhou}, {Tatematsu}, {Chibueze},
  {Su}, {Sun}, {Ristorcelli}, \& {Toth}}]{Xu2023ASSEMBLE}
{Xu}, F., {Wang}, K., {Liu}, T., {et~al.} 2023{\natexlab{a}}, arXiv e-prints,
  arXiv:2309.14684, \dodoi{10.48550/arXiv.2309.14684}

\bibitem[{{Xu} {et~al.}(2023{\natexlab{b}}){Xu}, {Wang}, {Liu}, {Goldsmith},
  {Zhang}, {Juvela}, {Liu}, {Qin}, {Li}, {Tej}, {Garay}, {Bronfman}, {Li},
  {Wu}, {G{\'o}mez}, {V{\'a}zquez-Semadeni}, {Tatematsu}, {Ren}, {Zhang},
  {Toth}, {Liu}, {Yue}, {Zhang}, {Baug}, {Issac}, {Stutz}, {Liu}, {Fuller},
  {Tang}, {Zhang}, {Dewangan}, {Lee}, {Zhou}, {Xie}, {Jiao}, {Wang}, {Liu},
  {Luo}, {Soam}, \& {Eswaraiah}}]{2023MNRAS.520.3259X}
{Xu}, F.-W., {Wang}, K., {Liu}, T., {et~al.} 2023{\natexlab{b}}, \mnras, 520,
  3259, \dodoi{10.1093/mnras/stad012}

\bibitem[{{Xu} {et~al.}(2016){Xu}, {Li}, {Zhang}, {Liu}, {Wang}, {Ning}, \&
  {Ju}}]{2016ApJ...819..117X}
{Xu}, J.-L., {Li}, D., {Zhang}, C.-P., {et~al.} 2016, \apj, 819, 117,
  \dodoi{10.3847/0004-637X/819/2/117}

\bibitem[{{Yang} {et~al.}(2018){Yang}, {Thompson}, {Urquhart}, \&
  {Tian}}]{2018ApJS..235....3Y}
{Yang}, A.~Y., {Thompson}, M.~A., {Urquhart}, J.~S., \& {Tian}, W.~W. 2018,
  \apjs, 235, 3, \dodoi{10.3847/1538-4365/aaa297}

\bibitem[{{Yang} {et~al.}(2022){Yang}, {Urquhart}, {Wyrowski}, {Thompson},
  {K{\"o}nig}, {Colombo}, {Menten}, {Duarte-Cabral}, {Schuller}, {Csengeri},
  {Eden}, {Barnes}, {Traficante}, {Bronfman}, {Sanchez-Monge}, {Ginsburg},
  {Cesaroni}, {Lee}, {Beuther}, {Medina}, {Mazumdar}, \&
  {Henning}}]{2022A&A...658A.160Y}
{Yang}, A.~Y., {Urquhart}, J.~S., {Wyrowski}, F., {et~al.} 2022, \aap, 658,
  A160, \dodoi{10.1051/0004-6361/202142039}

\bibitem[{{Yang} {et~al.}(2023){Yang}, {Dzib}, {Urquhart}, {Brunthaler},
  {Medina}, {Menten}, {Wyrowski}, {Ortiz-Le{\'o}n}, {Cotton}, {Gong}, {Dokara},
  {Rugel}, {Beuther}, {Pandian}, {Csengeri}, {Veena}, {Roy}, {Nguyen},
  {Winkel}, {Ott}, {Carrasco-Gonzalez}, {Khan}, \&
  {Cheema}}]{2023arXiv231009777Y}
{Yang}, A.~Y., {Dzib}, S.~A., {Urquhart}, J.~S., {et~al.} 2023, arXiv e-prints,
  arXiv:2310.09777, \dodoi{10.48550/arXiv.2310.09777}

\bibitem[{{Young} \& {Evans}(2005)}]{2005ApJ...627..293Y}
{Young}, C.~H., \& {Evans}, Neal~J., I. 2005, \apj, 627, 293,
  \dodoi{10.1086/430436}

\bibitem[{{Zhou} {et~al.}(2022){Zhou}, {Liu}, {Evans}, {Garay}, {Goldsmith},
  {G{\'o}mez}, {V{\'a}zquez-Semadeni}, {Liu}, {Stutz}, {Wang}, {Juvela}, {He},
  {Li}, {Bronfman}, {Liu}, {Xu}, {Tej}, {Dewangan}, {Li}, {Zhang}, {Zhang},
  {Ren}, {Tatematsu}, {Shing Li}, {Won Lee}, {Baug}, {Qin}, {Wu}, {Peng},
  {Zhang}, {Liu}, {Luo}, {Ge}, {Saha}, {Chakali}, {Zhang}, {Kim},
  {Ristorcelli}, {Shen}, \& {Li}}]{2022MNRAS.514.6038Z}
{Zhou}, J.-W., {Liu}, T., {Evans}, N.~J., {et~al.} 2022, \mnras, 514, 6038,
  \dodoi{10.1093/mnras/stac1735}

\end{thebibliography}
\bibliographystyle{aasjournal}

\end{document}